\documentclass[superscriptaddress, reprint, longbibliography,
 amsmath,amssymb,
aps,
prb,
]{revtex4-1}

\usepackage{simpler-wick}
\usepackage{graphicx}
\usepackage{dcolumn}
\usepackage{bm}
\usepackage{subfig}
\usepackage{feynmf}

\usepackage[colorlinks=true,linkcolor=blue,citecolor=blue,urlcolor=blue]{hyperref}
\usepackage{braket}

\DeclareMathOperator{\tr}{tr}

\DeclareMathOperator{\occ}{occ}

\usepackage{bbm}

\newcommand{\BCC}{\mathrm{BCC}}

\newcommand{\h}{\mathrm{h}}
\newcommand{\p}{\mathrm{p}}
\newcommand{\CC}{\mathrm{CC}}
\newcommand{\IP}{\mathrm{IP}}

\newcommand{\RPA}{\mathrm{RPA}}
\newcommand{\BSE}{\mathrm{BSE}}

\newcommand{\pt}{\mathrm{pt}}

\newcommand{\rf}{\mathrm{ref}}

\begin{document}
	
\title{Diagrammatic theory of the irreducible coupled-cluster self-energy}
	
\author{Christopher J. N. Coveney}
\email{christopher.coveney@physics.ox.ac.uk}
\affiliation{ Department of Physics, University of Oxford, Oxford OX1 3PJ, United Kingdom}
\author{David P. Tew}
\email{david.tew@chem.ox.ac.uk}
\affiliation{Physical and Theoretical Chemistry Laboratory, University of Oxford, South Parks Road, United Kingdom}

\date{\today}

\begin{abstract}
Coupled-cluster and Green's function theories are highly successful in treating many-body electron correlation, and there has been significant interest in identifying and leveraging connections between them. Here we present a diagrammatic definition of the irreducible coupled-cluster self-energy that directly embeds coupled-cluster (CC) theory within the framework of many-body field theory. The equation-of-motion coupled-cluster (EOM-CC) treatment emerges naturally from our definition via the Dyson equation and the Bethe-Salpeter equation (BSE), providing a unified description of the random phase approximation (RPA), $GW$-BSE, and CC theory for ground state and excitation energies. This clarifies the origin of previously established connections between RPA, $GW$-BSE, and CC theory, and it exposes the relationship between vertex corrections and the coupled-cluster amplitude equations.

\end{abstract}
	
	
\maketitle

\begin{fmffile}{diagram}
\section{Introduction}
The single-particle Green's function is one of the central objects in the quantum theory of many-particle systems.~\cite{Quantum,mahan2000many,stefanucci2013nonequilibrium} From knowledge of the single-particle Green's function we can obtain the exact addition and removal energies, the expectation value of any single-particle observable and, in the case of a two-body interaction, the exact ground state energy of a system of many interacting particles. In many-body field theory, the irreducible self-energy has equal importance.~\cite{dyson1949radiation} The self-energy connects a non-interacting to an interacting Green's function through the Dyson equation and implicitly contains all the many-body interactions a single-particle experiences. The usefulness of the self-energy stems from the fact that it may be defined in terms of a Feynman diagram series expansion without \emph{a priori} knowledge of the Green's function. 

Hedin’s equations define a formally exact procedure for generating the many-body self-energy by expressing its infinite Feynman diagram series in terms of the vertex function.~\cite{hedin1965new} However, no known functional form for the vertex exists and inclusion of vertex corrections to access higher-order many-body correlation effects is neither rigorous nor trivial.~\cite{pokhilko2021evaluation,martin2016interacting,verdozzi1995evaluation,von2009successes,romaniello2009self,von2010kadanoff,romaniello2012beyond} The lack of a controllable and systematically improvable expression for the self-energy in traditional field theory arises due to the fact that these approaches do not assume a closed form expression for the exact many-body ground state wavefunction.~\cite{kohn1960ground,luttinger1960ground,luttinger1961analytic,potthoff2014making,van2013gw} 

Ground state coupled-cluster (CC) theory represents a reformation of the Configuration Interaction that preserves the correct many-body product structure of the electronic ground state wavefunction. As a result, CC theory allows for proper parameterisation of $N$-body correlation effects that lead to a hierarchical, systematically improvable set of equations for the exact ground state energy and wavefunction. Commonly, ionization potentials, electron affinities and neutral excitation energies in coupled-cluster theory are obtained from application of linear response equation-of-motion theory to the coupled-cluster ansatz (EOM-CC). This is in contrast to the field theoretic approach which is not based on a wavefunction representation but instead works with dynamical correlation functions known as many-body Green's functions. 

Nooijen and Snijders' formulation of the coupled-cluster Green's function (CCGF) was based on the direct construction of the Green's function using quantities obtained from ground state and EOM-CC theory and has led to many practical advantages over conventional many-body theory for the computation of addition and removal energies.~\cite{nooijen1992coupled,nooijen1993coupled,peng2018green,shee2019coupled,backhouse2022constructing} However, this formulation does not involve the explicit construction of the irreducible coupled-cluster self-energy and is not yet fully connected to the conventional formulation of many-body field theory. As a result, the connections between the CCGF and the CC Bethe-Salpeter equation have remained elusive.

The recent analysis of the supermatrix formulation of the quasiparticle equation within the $GW$ and GF2 approximations has led to significant advances in our understanding of the dynamical effects contained in the many-body self-energy.~\cite{bintrim2021full,bintrim2022full,scott2023moment,quintero2022connections,monino2023connections,tolle2023exact} The relationship between the random-phase approximation (RPA) and the coupled-cluster doubles (CCD) amplitude equations by retaining only `ring coupled-cluster' diagrams is now well established.~\cite{scuseria2008ground,berkelbach2018communication,rishi2020route,quintero2022connections} Recent work on uncovering formal and numerical similarities between the $G_0W_0$ supermatrix and \emph{approximate} forms of the ionization potential/electron affinity EOM-CC singles and doubles (IP/EA-EOM-CCSD) eigenvalue problem has also demonstrated how improved formulations of the quasiparticle equation may be constructed.~\cite{lange2018relation,tolle2023exact}

However, no diagrammatic definition of the coupled-cluster self-energy has been introduced that makes clear its relationship to conventional Green's function theory. This is the subject of the present paper. Drawing on the preceding work, here we reveal the structure of the coupled-cluster self-energy along with its relationship to EOM-CC and the Green's function formalism. This allows us to define the notion of a coupled-cluster
self-energy that retains the functional role of the self-energy in conventional
many-body theory in how it is used to define equations governing
different observables such as charged and neutral excitation energies. The power of this approach rests in the formally exact coupled-cluster ansatz for the ground state wavefunction and a self-energy that directly connects to that of conventional many-body field theory. 

The paper is organized as follows. In Section~\ref{sec:eom_cc}, we provide an overview of the single-particle Green's function, self-energy and coupled-cluster theory. In Section~\ref{sec:sim_mbgf}, we derive the relationship between the effective interactions introduced by the similarity transformed Hamiltonian and many-body Green's functions to derive the coupled-cluster self-energy of the coupled-cluster Lagrangian. In Section~\ref{sec:BSE}, we define the coupled-cluster BSE kernel from the coupled-cluster self-energy, thereby demonstrating the origin of the connections between RPA, $GW$-BSE and coupled-cluster theory for neutral excitation energies. In Section~\ref{sec:rel_eomcc}, we explicitly outline the relationship between EOM-CC theory and the coupled-cluster self-energy for charged excitations. Finally, we summarize our findings and provide outlook for future work in Section~\ref{sec:conc}.

\section{single-particle Green's function and Coupled-cluster theory}~\label{sec:eom_cc}

In this section we present a brief overview of Green's function theory, ground state coupled-cluster theory and equation-of-motion coupled-cluster theory. In the following, indices and wavevectors $i\mathbf{k}_i,j\mathbf{k}_j,\cdots$ denote occupied (valence band) spin-orbitals, $a\mathbf{k}_a,b\mathbf{k}_b,\cdots$ virtuals (conduction band) and $p\mathbf{k}_p,q\mathbf{k}_q,\cdots$ general spin-orbitals. The wavevector dependence is omitted for clarity. 

\subsection{Electronic Green's function and self-energy}
The single-particle electronic Green's function is defined as
\begin{equation}~\label{eq:gf}
    iG_{pq}(t_1-t_2) = \braket{\Psi_0|\mathcal{T}\left\{a_{p}(t_1)a^\dag_{q}(t_2)\right\}|\Psi_0} ,
\end{equation}
where $a^\dag_p(t)/a_p(t)$ create/annihilate electrons in spin-orbital $\ket{\phi_p}$ in the Heisenberg representation, $\mathcal{T}$ is the time-ordering operator and $\ket{\Psi_0}$ is the normalized, exact ground state.~\cite{Quantum} The poles of the exact Green's function correspond to the exact ionization potentials and electron affinities of the system. From the equation of motion of the Green's function, the exact ground state energy can be obtained from the Galitskii-Migdal formula.~\cite{galitskii1958application} Therefore, the Green's function simultaneously contains both the exact single-particle spectrum and ground state energy. From the Dyson equation, it can be shown that the ionization potentials, electron affinities and Dyson orbitals can be obtained by solution of the quasiparticle equation~\cite{hybertsen1986electron,Quantum} 
\begin{equation}\label{eq:qp_Eq.}
    [f + \Sigma^c(\omega=\varepsilon_p)]\ket{\psi_{p}} = \varepsilon_{p}\ket{\psi_p}  .
\end{equation}
Here, $f$ is the Fock operator and $\Sigma^c$ is the many-body self-energy. In this work, we choose to partition the electronic self-energy with respect to the reference Fock operator. However, the electronic self-energy can be constructed and partitioned with respect to any general `mean field' reference operator.~\cite{hedin1965new,hybertsen1986electron} The frequency-dependence of the self-energy captures the higher-order interactions whilst remaining in the single-particle spin-orbital basis. In contradistinction with the analytic diagrammatic expansion, the exact self-energy may also be defined by direct inversion of the Dyson equation
\begin{gather}
    \begin{split}~\label{eq:dys}
        \Sigma^c_{pq}(\omega) = [G^{0}_{pq}(\omega)]^{-1}-G^{-1}_{pq}(\omega) \ ,
    \end{split}
\end{gather}
where $G^{0}_{pq}$ is the `non-interacting' Green's function obtained by replacing $\ket{\Psi_0}$ in Eq.~\ref{eq:gf} with the reference, $\ket{\Phi_0}$, where the time-dependence of the field operators is governed by the Fock operator. This is not a functional relationship as it requires \emph{a priori} knowledge of the exact Green's function; the power of the Dyson equation turns on the separate existence of a diagrammatic expansion for the self-energy which preserves the analytic pole structure of the Green's function. 

The spectral form of the electronic many-body self-energy is given by~\cite{schirmer1983new,caruso2013self,raimondi2018algebraic,schirmer2018many}
\begin{gather}
    \begin{split}~\label{eq:spec_se}
        \Sigma^c_{pq}(\omega) &= \Sigma^{\infty}_{pq} + \sum_{JJ'} U^\dag_{pJ}\left[(\omega+i\eta)\mathbbm{1}-(\mathbf{K}^{>}+\mathbf{C}^{>})\right]^{-1}_{JJ'}U_{J'q}\\
        &+ \sum_{AA'} V_{pA}\left[(\omega-i\eta)\mathbbm{1}-(\mathbf{K}^{<}+\mathbf{C}^{<})\right]^{-1}_{AA'}V^\dag_{A'q} \ .
    \end{split}
\end{gather}
Using this expression, the frequency-dependent quasiparticle equation (Eq.~\ref{eq:qp_Eq.}) is equivalent to diagonalization of the Dyson supermatrix~\cite{raimondi2018algebraic,schirmer2018many}
\begin{gather}
    \begin{split}~\label{eq:el_dys}
        \mathbf{D} = \left(\begin{array}{ccc}
            f_{pq}+\Sigma^{\infty}_{pq} & U^\dag_{pJ'} & V_{pA'} \\
            U_{Jq} & (\mathbf{K}^{>}_{JJ'}+\mathbf{C}^{>}_{JJ'}) & \mathbf{0} \\
             V^\dag_{Aq} & \mathbf{0} & (\mathbf{K}^{<}_{AA'}+\mathbf{C}^{<}_{AA'}) 
        \end{array}\right) .
    \end{split}
\end{gather}
The indices $JJ'/AA'$ label forward-time/backward-time Intermediate State Configurations (ISCs) and outline the character of the different multi-particle-hole configurations that a single-particle state may be connected to. ISCs are excited state configurations containing $(N\pm1)$-electrons that mediate between exact energy eigenstates and electronic configurations with respect to the reference.~\cite{schirmer2018many} The matrices $(\mathbf{K}^{>}_{JJ'}+\mathbf{C}^{>}_{JJ'})$ and $(\mathbf{K}^{<}_{AA'}+\mathbf{C}^{<}_{AA'})$ represent the interactions between the different ISCs. The quantities $U_{qJ}$ and $V_{pA}$ represent the coupling matrices that link initial and final single-particle states to the ISCs. 
The upper left block of $\mathbf{D}$ is defined over the complete set of single-particle spin-orbitals, with the coupling and interaction matrices containing coupling to all possible ISCs. The zero entries of the Dyson supermatrix are present as the forward and backward time self-energy contributions are coupled only through the initial and final single-particle states. Diagonalization of the Dyson supermatrix yields the complete set of ionization potentials and electron affinities. 

From knowledge of the electronic self-energy and single-particle Green's function, the ground state correlation energy may be found from the Galitskii-Migdal formula as~\cite{galitskii1958application,coveney2023regularized,backhouse2020efficient}
\begin{gather}
    \begin{split}~\label{eq:GM}
        E_c &= -\frac{i}{2}\lim_{\eta\to0^+}\sum_{pq}\int^{\infty}_{-\infty}\frac{d\omega}{2\pi} e^{i\eta\omega}\Sigma^{c}_{pq}(\omega)G_{qp}(\omega) \\
        &= -\frac{i}{2} \int d3\ \Sigma(1,3)G(3,1^+) \ ,
    \end{split}
\end{gather}
where the second equality stems from the space-spin-time representation, $1 \equiv (\mathbf{r}_1\sigma_1,t_1)$ is the composite spin-space-time coordinate and $\int d1 \equiv \int d\mathbf{r}_1d\sigma_1\int^{\infty}_{-\infty}dt_1$.
At zero temperature, the self-energy is given by the functional derivative of the ground state correlation energy with respect to the Green’s function~\cite{luttinger1960ground,luttinger1961analytic}
\begin{gather}
    \begin{split}~\label{eq:LW}
        \Sigma^{c}(1,2) = i\frac{\delta E_c}{\delta G(2,1)}\ . 
    \end{split}
\end{gather}
Eq.~\ref{eq:LW} highlights the fact that the self-energy may be constructed as a functional of the dressed Green’s function and two-body coulomb interaction.~\cite{stefanucci2013nonequilibrium,hedin1965new} Finally, the relationship between the one-body reduced density matrix and the single-particle Green's function is given by
\begin{gather}
    \begin{split}
        G_{pq}(t-t^{+}) = i\braket{\Psi_0|a^\dag_qa_{p}|\Psi_0} \ ,
    \end{split}
\end{gather}
where the notation, $G_{pq}(t-t^{+})$, denotes evaluation of the second time argument of the Green's function with a positive infinitesimal, \emph{i.e.} $t^{+} = t+\eta$, where $\eta\to0$ from above. 

\subsection{Overview of coupled-cluster theory}
In CC theory, the many-body electronic ground state is expressed as 
\begin{subequations}
\begin{gather}
    \begin{split}
        \ket{\Psi^\CC_0}=e^{T}\ket{\Phi} ,
    \end{split}
\end{gather}
where $T$ creates all excitations with respect to a reference determinant $\ket{\Phi}$ as
    \begin{gather}
\begin{split}
        T &= \sum_{ai} t^{a}_i a^\dag_aa_i + \frac{1}{4} \sum_{abij} t^{ab}_{ij} a^\dag_aa^\dag_ba_ja_i \\
    &+ \frac{1}{36} \sum_{abcijk} t^{abc}_{ijk} a^\dag_aa^\dag_ba^\dag_c a_k a_ja_i + \cdots
\end{split}
\end{gather}
\end{subequations}
Here, $\{t^{a}_i,t^{ab}_{ij},t^{abc}_{ijk},\cdots\}$ are the cluster amplitudes. The ground state energy and cluster amplitudes are obtained by projection onto the manifold of Slater determinants
\begin{subequations}
    \begin{gather}
        \begin{split}
            \braket{\Phi|\Bar{H}|\Phi} = E^\CC_0
        \end{split}
    \end{gather}
    \begin{gather}
        \begin{split}
            \braket{\Phi^{a}_{i}|\Bar{H}|\Phi} = 0
        \end{split}
    \end{gather}
    \begin{gather}
        \begin{split}
        \braket{\Phi^{ab}_{ij}|\Bar{H}|\Phi} = 0
        \end{split}
    \end{gather}
    and so on, where 
    \begin{equation}
        \Bar{H} = e^{-T}He^{T}
    \end{equation}
\end{subequations}
is the similarity transformed Hamiltonian. The  manifold of excited Slater determinants are given by $\ket{\Phi^a_i} = a^\dag_aa_i\ket{\Phi}$ etc. When $T$ is not truncated, the ground state energy obtained from this procedure is formally exact. Importantly, the similarity transformed Hamiltonian $\Bar{H}$, is non-Hermitian, possessing different left and right eigenstates. 

In standard coupled-cluster theory, the reference determinant, $\ket{\Phi}$, corresponds to the Hartree--Fock wavefunction constructed from spin-orbitals that are determined by separate optimization of the reference. However, coupled-cluster theory may also be expressed with respect to an arbitrary Slater determinant in an alternative spin-orbital basis. One of the most useful reformulations of coupled-cluster theory is Brueckner coupled-cluster (BCC), whereby the spin-orbitals are determined simultaneously with the optimization of the cluster amplitudes.~\cite{scuseria1995connections,helgaker2013molecular} This is performed by an alternative parametrization of the singles amplitudes by rotating the spin-orbital basis via 
\begin{gather}
\begin{split}
    \ket{\Phi_0}=\exp\left(-\kappa\right)\ket{\Phi} ,
\end{split}
\end{gather}
where 
\begin{gather}
    \begin{split}
        \kappa=\sum_{ai}\kappa_{ai}\left(a^\dag_a a_i-a^\dag_i a_a\right)
    \end{split}
\end{gather}
is anti-Hermitian $\kappa=-\kappa^\dag$, and $\kappa_{ai}$ is the orbital-rotation matrix to be determined. The orbital-rotation results in the corresponding unitary transformation of the creation (annihilation) operators
\begin{gather}
    \begin{split}
        \tilde{a}^\dag_p = e^{-\kappa}a^\dag_pe^{\kappa} 
    \end{split}
\end{gather}
to the new spin-orbital basis. Within Brueckner theory, the orbital-rotation parameters, $\{\kappa_{ai}\}$, are determined by the singles amplitude equations. The result is that the BCC ground state wavefunction has the form of a standard CC wavefunction, but where the singles amplitudes, $\{t^{a}_{i}\}$, are eliminated by construction as they are already `accounted' for by transformation to the Brueckner basis. For simplicity, in the rest of this paper we will work in the Brueckner basis where we have already accounted for the orbital-rotation. However, our formalism can be easily generalized to include singles amplitudes.

As first proposed in Refs~\citenum{nooijen1992coupled} and~\citenum{nooijen1993coupled}, one may construct the CCGF from the corresponding EOM-CC eigenvalue problem. The IPs and EAs are the eigenvalues of the similarity transformed Hamiltonian $\Bar{H}$ in the basis of all Slater determinants containing ($N\pm1$)-electrons. For example, the IP-EOM-CC eigenvalue problem is written for the right eigenstates as
\begin{equation}~\label{eq:eom_cc}
        -\Bar{H}_NR^{\IP}_{\nu}\ket{\Phi_0} = \varepsilon^{N-1}_{\nu}R^{\IP}_{\nu}\ket{\Phi_0} , 
\end{equation}
where $\Bar{H}_N=\Bar{H}-E^{\CC}_0$ is the normal-ordered similarity transformed Hamiltonian and the state $R^{\IP}_{\nu}\ket{\Phi_0}$ is the exact $N-1$ electron, right eigenstate of the similarity transformed Hamiltonian. Importantly, Eq.~\ref{eq:eom_cc} implicitly requires $[R^{\IP}_{\nu},T]=0$. When $T$ is not truncated, the eigenvalues $\varepsilon_{\nu}^{N-1}$ are the exact ionization potentials of the system and correspond to the poles of the exact Green's function.~\cite{nooijen1992coupled,nooijen1993coupled,lange2018relation} In IP-EOM-CC theory, the equation-of-motion operator creates all determinants consisting of $(N-1)$-electrons and corresponds to the choice
\begin{equation}~\label{eq:eom}
    R^{\IP}_{\nu} = \sum_{i} r_{i}(\nu)a_{i} + \sum_{i<j,a}r^{a}_{ij}(\nu)a^\dag_aa_ja_i + \cdots
\end{equation}
With this definition of the equation-of-motion operator, Eq.~\ref{eq:eom_cc} is re-written in terms the IP-EOM-CC supermatrix
\begin{gather}
    \begin{split}~\label{eq:eom_bcc}
       -\mathbf{\bar{H}}&^{\text{IP-EOM-CC}} =\\
       & \left(\begin{array}{ccc}
           \braket{\Phi_{i}|\bar{H}_N|\Phi_{j}}  & \braket{\Phi_{i}|\bar{H}_N|\Phi^{a}_{kl}} & \cdots \\
            \braket{\Phi^{b}_{mn}|\bar{H}_N|\Phi_{j}} & \braket{\Phi^{b}_{mn}|\bar{H}_N|\Phi^{a}_{kl}} & \cdots \\
             \vdots & \vdots & \ddots
        \end{array}\right) 
    \end{split}
\end{gather}
Diagonalization of this supermatrix yields the left and right $(N-1)$-electron eigenstates of $\bar{H}$. The left and right eigenstates of the IP/EA-EOM-CC eigenvalue problem form a complete biorthogonal set in the $(N\pm1)$-electron space, and can be used to construct the CCGF from the Lehmann representation. However, the associated coupled-cluster self-energy exists only numerically by explicit inversion of the Dyson equation (Eq.~\ref{eq:dys}) and does not admit a diagrammatic definition.~\cite{nooijen1992coupled,nooijen1993coupled} Thus, this approach to the coupled-cluster self-energy does not expose its relationship to the Bethe-Salpeter kernel and the general formalism of many-body field theory. Additionally, calculation of the ground state energy at any level of approximation for the CCGF results in a different ground state energy from that obtained at the same level of approximation of the ground state coupled-cluster equations.~\cite{nooijen1992coupled,nooijen1993coupled} 

In this work, we introduce an alternative formalism that focuses on constructing the coupled-cluster self-energy based on the universal relationships between the interaction kernels of many-body field theory.~\cite{hedin1965new,kohn1960ground,luttinger1960ground}

\section{Relationship between the similarity transformed Hamiltonian and many-body Green's functions}~\label{sec:sim_mbgf}

To expose the relationship between the effective interactions generated by the similarity transformed Hamiltonian and many-body Green's functions, it is necessary to outline the relationship between normal-ordering and Green's functions.

\subsection{Relationship between the Fock operator and `non-interacting' Green's function}
The electronic structure Hamiltonian is written in second quantization as 
\begin{equation}
    \begin{split}
        H = \sum_{pq} h_{pq} a^\dag_pa_q + \frac{1}{4}\sum_{pq,rs} \braket{pq||rs}a^\dag_pa^\dag_qa_sa_r
    \end{split}
\end{equation}
where $h_{pq}$ is the one-body term that includes the kinetic energy and attractive interaction due to the potential of the nuclei, whilst $\braket{pq||rs}=\braket{pq|rs}-\braket{pq|sr}$ is the antisymmetrized Coulomb interaction. The electronic structure Hamiltonian may be normal-ordered with respect to a reference determinant to give 
\begin{equation}
    \begin{split}
        H = E^{\rf}_0 + \sum_{pq} f_{pq}\{a^\dag_{p}a_q\}_0 +\frac{1}{4}\sum_{pq,rs} \braket{pq||rs}\{a^\dag_{p}a^\dag_qa_{s}a_r\}_0
    \end{split}
\end{equation}
where the notation $\{\cdots\}_0$ indicates a normal-ordered string of operators with respect to the reference, $\ket{\Phi_0}$. The reference energy is given by the expectation value of the Hamiltonian over the reference state, $E^{\rf}_0 = \braket{\Phi_0|H|\Phi_0}$. The Fock operator, $f_{pq}$, is generated by the procedure of normal-ordering by contraction of the Coulomb interaction with the `non-interacting' one-body reduced density matrix via 
\begin{equation}~\label{eq:fok}
    f_{pq} = h_{pq}-i\sum_{rs}\braket{pr||qs}G^{0}_{sr}(t-t^+) \ .
\end{equation}
In what follows, we find it useful to introduce the following notation
\begin{gather}
    \begin{split}~\label{eq:equal_t}
        \tilde{G}_{pq} = G^{0}_{pq}(t-t^{+}) = i\braket{\Phi_0|a^\dag_qa_{p}|\Phi_0}
    \end{split}
\end{gather}
to denote the equal-time `non-interacting' Green's function. Therefore, we re-write the Fock operator as 
\begin{equation}
    f_{pq} = h_{pq}-i\sum_{rs}\braket{pr||qs}\tilde{G}_{sr} \ ,
\end{equation}
with the property that 
\begin{gather}
    \begin{split}~\label{eq:fokk}
        i\frac{\delta f_{pr}}{\delta\tilde{G}_{sq}} = \braket{pq||rs} \ .
    \end{split}
\end{gather}

\subsection{Normal-ordering the similarity transformed Hamiltonian}

The similarity transformed Hamiltonian, in its full generality, is an $N$-body operator when $T$ is not truncated and is written as
\begin{gather}
\begin{split}
    \Bar{H} &= \sum_{pq}\bar{h}_{pq}a^\dag_{p}a_q + \frac{1}{4}\sum_{pq,rs} \bar{h}_{pq,rs}a^\dag_{p}a^\dag_qa_{s}a_r\\
    &+\frac{1}{36}\sum_{pqr,stu}\bar{h}_{pqr,stu} a^\dag_{p}a^\dag_q a^\dag_{r}a_ua_ta_s + \cdots \ ,
\end{split}
\end{gather}
where $\{\bar{h}_{pq,rs},\bar{h}_{pqr,stu},\cdots\}$ are antisymmetrized two-body, three-body and so on, matrix elements. Normal-ordering $\bar{H}$ with respect to the reference determinant gives
\begin{gather}
\begin{split}~\label{eq:norm_sim}
    \Bar{H} &= E^{\CC}_0 + \sum_{pq}F_{pq}\{a^\dag_{p}a_q\}_0 + \frac{1}{4}\sum_{pq,rs} \chi_{pq,rs}\{a^\dag_{p}a^\dag_qa_{s}a_r\}_0\\
    &+\frac{1}{36}\sum_{pqr,stu}\chi_{pqr,stu} \{a^\dag_{p}a^\dag_q a^\dag_{r}a_ua_ta_s\}_0 + \cdots
\end{split}
\end{gather}
where $\{F_{pq},\chi_{pq,rs},\chi_{pqr,stu},\cdots\}$ are the one-body, antisymmetrized two-body, three-body and so on, effective interaction matrix elements.~\cite{shavitt2009many}
The explicit expressions for the effective interaction matrix elements, $\{F_{pq},\chi_{pq,rs},\chi_{pqr,stu},\cdots\}$, up to the four-body interaction elements can be found in Refs~\citenum{shavitt2009many} and~\citenum{gauss1995coupled}. The amplitude equations are contained in the effective interaction elements as 
\begin{subequations}
    \begin{gather}
        \begin{split}~\label{eq:amp1}
            F_{ai} = \braket{\Phi^{a}_i|\Bar{H}_N|\Phi_0}
        \end{split}
    \end{gather}
    \begin{gather}
        \begin{split}~\label{eq:amp2}
           \chi_{ab,ij} = \braket{\Phi^{ab}_{ij}|\Bar{H}_N|\Phi_0}
        \end{split}
    \end{gather}
    \begin{gather}
        \begin{split}~\label{eq:amp3}
            \chi_{abc,ijk} = \braket{\Phi^{abc}_{ijk}|\Bar{H}_N|\Phi_0} 
        \end{split}
    \end{gather}
\end{subequations}
and so on, which all evaluate to zero at convergence.~\cite{shavitt2009many} 
For what follows, it is central to note that the following relationships hold (see Appendix~\ref{app:1})
\begin{subequations}
    \begin{gather}
    \begin{split}~\label{eq:proof}
        \chi_{pq,rs} = i\frac{\delta F_{pr}}{\delta\tilde{G}_{sq}} \ ,
    \end{split}
\end{gather}
\begin{gather}
    \begin{split}~\label{eq:proof1}
        \chi_{pqr,stu} = i\frac{\delta \chi_{pq,st}}{\delta\tilde{G}_{ur}} \ ,
    \end{split}
\end{gather}
\end{subequations}
and so on for the higher-order effective interactions appearing in the normal-ordered similarity transformed Hamiltonian. Eqs~\ref{eq:proof} and~\ref{eq:proof1} and so on for the higher-order effective interactions can also be viewed as generalizations of Eq.~\ref{eq:fokk} for the relationship between the Fock operator and the two-body Coulomb interaction.

By the normal-ordering procedure, the corresponding effective interaction matrix elements arise by contraction with density matrices which are related to many-body Green's functions as follows
\begin{subequations}
    \begin{gather}
        \begin{split}~\label{eq:eff1}
            F_{pq} &= \bar{h}_{pq} -i \sum_{rs} \bar{h}_{pr,qs}G^{0}_{sr}(t-t^+) \\
            &+ \frac{i}{4}\sum_{rs,tu} \bar{h}_{prs,qtu}G^{2\p\h,(0)}_{tu,rs}(t-t^+) + \cdots
        \end{split}
    \end{gather}
    \begin{gather}
        \begin{split}~\label{eq:eff2}
            \chi_{pq,rs} &= \bar{h}_{pq,rs} -i\sum_{tu} \bar{h}_{pqt,rsu}G^0_{ut}(t-t^+) \\
            &+ \frac{i}{4}\sum_{tuwv}\bar{h}_{pqtu,rswv}G^{2\p\h,(0)}_{wv,tu}(t-t^+) + \cdots
        \end{split}
    \end{gather}
    \begin{gather}
        \begin{split}~\label{eq:eff3}
            \chi_{pqr,stu} &= \bar{h}_{pqr,stu} -i\sum_{wv} \bar{h}_{pqrw,stuv}G^0_{vw}(t-t^+) \\
            &+ \frac{i}{4}\sum_{wvo\sigma}\bar{h}_{pqrwv,stuo\sigma}G^{2\p\h,(0)}_{o\sigma,wv}(t-t^+) + \cdots
        \end{split}
    \end{gather}
\end{subequations}
    and so on, where $G^{2\p\h,(0)}$ is the non-interacting two-particle-hole Green's function defined in Appendix~\ref{app:1}. Implicitly, all expressions for the effective interaction matrix elements continue with the contraction of the higher-order matrix elements with the corresponding higher-order Green's functions. These expressions represent a generalization of Eq.~\ref{eq:fok} in the presence of higher-body interactions.

\subsection{The coupled-cluster self-energy of the Lagrangian}

The BCC energy is given by the extremum of the Lagrangian equation
\begin{equation}\label{eq:BCC_energy}
	\mathcal{L}^{\BCC}_c = \braket{\Phi_0|\Bar{H}_N|\Phi_0}  + \sum_{\mu}\lambda_{\mu}\braket{\Phi_{\mu}|\Bar{H}_N|\Phi_0}\ , 
\end{equation}
where $\braket{\Phi_{\mu}|\Bar{H}_N|\Phi_0}$ represent the complete set of amplitude equations, with $\bar{H}_N = \bar{H}-E^{\rf}_0$, and $\lambda_{\mu}=\{\lambda^{i}_{a},\lambda^{ij}_{ab},\lambda^{ijk}_{abc},\cdots\}$ are the associated Lagrange multipliers used to determine the left ground eigenstate of $\Bar{H}$. The first term on the right hand side of Eq.~\ref{eq:BCC_energy} is the Brueckner correlation energy, $E^{\BCC}_c$. At convergence, the second term of the right hand side of Eq.~\ref{eq:BCC_energy} vanishes as the amplitude equations are satisfied, $\braket{\Phi_\mu|\Bar{H}_N|\Phi_0}=0$. The ground state correlation energy is expanded to give~\cite{shavitt2009many} 
\begin{gather}
\begin{split}~\label{eq:contract}
        E^{\BCC}_c = \frac{1}{16}\sum_{\substack{ijab\\pqrs}}&\braket{pq||rs}t^{ab}_{ij}\\
        &\times\braket{\Phi_0|\{a^\dag_pa^\dag_qa_sa_r\}_0\{a^\dag_aa^\dag_ba_ja_i\}_0|\Phi_0}  .
\end{split}
\end{gather}
Using Wick's theorem to evaluate this expression results in four equal contributions from the corresponding contractions to give the ground state correlation energy
\begin{gather}
\begin{split}
        E^{\BCC}_c = \frac{1}{4}\sum_{ijab}&\braket{ij||ab}t^{ab}_{ij} \ .
\end{split}
\end{gather}
\begin{figure*}[ht]
   \centering
    \includegraphics[width=130mm,height=17mm]{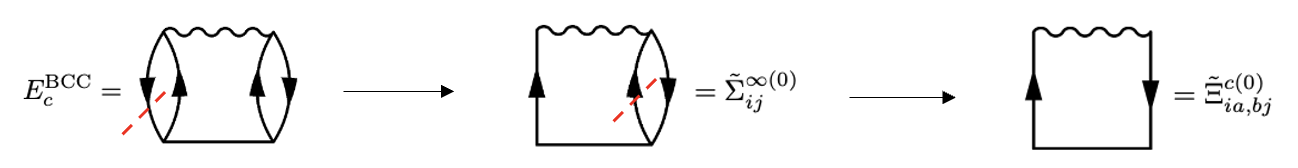}
    \caption{The series of coupled-cluster functional derivatives obtained by cutting lines in the Goldstone diagrams.}
    \label{fig:SE}
\end{figure*}
Clearly, the doubles amplitudes $t^{ab}_{ij}$, depend on all higher order amplitudes due to the structure of the CC amplitude equations. In the Brueckner basis, the reference energy is given by~\cite{scuseria1995connections}
\begin{gather}
    \begin{split}~\label{eq:energy_ref}
        E^{\rf}_0 &= \braket{\Phi_0|H|\Phi_0}=\sum_{i}h_{ii}+\frac{1}{2}\sum_{ij}\braket{ij||ij}\\
        &=\frac{1}{2}\sum_{i}(h_{ii}+f_{ii})\\
        &= -\frac{i}{2}\sum_{pq}(h_{pq}+f_{pq})\tilde{G}_{qp} \ . \\
    \end{split}
\end{gather}
Using the definition of the effective interaction matrix elements defined in Eq.~\ref{eq:eff1}, the Brueckner correlation energy is given by~\cite{scuseria1995connections,shavitt2009many}
\begin{gather}
    \begin{split}~\label{eq:BCC_gf}
        E^{\BCC}_{c}
        &=\frac{1}{2}\sum_{i}(F_{ii}-f_{ii}) = \frac{1}{4}\sum_{ijab} \braket{ij||ab}t^{ab}_{ij} \\
        &= -\frac{i}{2}\sum_{pq} (F_{pq}-f_{pq})\tilde{G}_{qp} .
    \end{split}
\end{gather}
Diagrammatically, Eq.~\ref{eq:BCC_gf} may be written as
\begin{gather}~\label{eq:feyn_diag}
\begin{split}E^{\BCC}_c &=\hspace{2.5mm}
\begin{gathered}
\begin{fmfgraph*}(40,40)
\fmfset{arrow_len}{3mm}
    \fmfleft{i1,i2}
    \fmfright{o1,o2}
    \fmf{fermion,left=0.3}{i1,i2}
    \fmf{fermion,left=0.3}{i2,i1}
    \fmf{plain}{i1,o1}
    \fmf{wiggly}{i2,o2}
    \fmf{fermion,left=0.3}{o1,o2}
    \fmf{fermion,left=0.3}{o2,o1}
\end{fmfgraph*}
\end{gathered}
\hspace{2.5mm}= \frac{1}{2}\times\hspace{10mm}
\begin{gathered}
    \begin{fmfgraph*}(40,40)
            \fmfset{arrow_len}{3mm}
           \fmfleft{i1}
            \fmfright{o1}
            \fmf{fermion,right=0.6}{i1,i1}
            \fmf{dbl_dashes}{o1,i1}
            \fmfdot{i1}
            \fmfv{decor.shape=cross,decor.filled=full, decor.size=1.5thic}{o1}
       \end{fmfgraph*}
   \end{gathered}\hspace{2.5mm},
\end{split}
\end{gather}
where the first diagram of Eq.~\ref{eq:feyn_diag} is interpreted as the Goldstone diagram of the coupled-cluster ground state correlation energy. The second diagram is a Feynman diagram with the arrow given by the equal-time `non-interacting' Green's function and the interaction line as
\begin{gather}
\begin{split}
  \begin{gathered}
\begin{fmfgraph*}(40,40)
    \fmfset{arrow_len}{3mm}
    \fmfleft{i1,i2,i3}
    \fmfright{o1,o2,o3}
    \fmf{fermion,label=$q$,label.side=left}{i1,i2}
    \fmf{fermion,label=$p$,label.side=left}{i2,i3}
    \fmf{dbl_dashes}{i2,o2}
    \fmfforce{(0.0w,0.h)}{i1}
    \fmfforce{(0.0w,0.5h)}{i2}
    \fmfforce{(0.0w,1.0h)}{i3}
    \fmfdot{i2}
    \fmfv{decor.shape=cross,decor.filled=full, decor.size=1.5thic}{o2}
\end{fmfgraph*}
\end{gathered}\hspace{5mm}= F_{pq}-f_{pq} \ .
\end{split}
\end{gather}
By the Feynman rules,~\cite{Quantum} the Feynman diagram corresponds exactly to Eq.~\ref{eq:BCC_gf} for the Brueckner correlation energy. 

Within Green's function theory, the Fock operator is given by the functional derivative of the reference energy with respect to the `non-interacting' Green's function as
\begin{equation}~\label{eq:conv_se}
	f_{pq} = i\frac{\delta E^{\rf}_0}{\delta \tilde{G}_{qp}} .
\end{equation}
Using Eqs~\ref{eq:fokk} and~\ref{eq:energy_ref}, we find that 
\begin{gather}
\begin{split}
        i\frac{\delta E^{\rf}_0}{\delta \tilde{G}_{qp}} &= \frac{1}{2}(h_{pq}+f_{pq}) -\frac{i}{2}\sum_{rs}\braket{pr||qs}\tilde{G}_{sr}\\
        &= h_{pq} -i\sum_{rs}\braket{pr||qs}\tilde{G}_{sr} = f_{pq}
\end{split}
\end{gather}
Therefore, the self-energy of the reference state is given by 
\begin{gather}
\begin{split}~\label{eq:se_ref}
        \Sigma^{\rf}_{pq} &=f_{pq}-h_{pq} \\
        &=-i\sum_{rs}\braket{pr||qs}\tilde{G}_{sr} =\sum_{i}\braket{pi||qi} \ .
\end{split}
\end{gather}

Within coupled-cluster theory, the exact ground state energy is expressed in terms of the non-interacting Green's function as it is the interaction vertices that are renormalized by introduction of the similarity transformed Hamiltonian, $\Bar{H}$. 
This is due to the fact that the right ground eigenstate of $\bar{H}$ is the reference determinant, $\ket{\Phi_0}$. Therefore, the natural generalization of Eq.~\ref{eq:se_ref} to include the effects of the CC similarity transformation is to take the functional derivative of the coupled-cluster Lagrangian with respect to the `non-interacting' Green's function. We define the coupled-cluster self-energy of the coupled-cluster Lagrangian as 
\begin{equation}\label{eq:occ_se}
    \Tilde{\Sigma}^{\infty}_{pq} =  i\frac{\delta \mathcal{L}^{\BCC}_{c}[\Tilde{G}]}{\delta\Tilde{G}_{qp}}\ ,
\end{equation}
where $\Tilde{G}_{qp}$ is the equal-time, `non-interacting' Green's function in the basis of the Brueckner orbitals defined in Eq.~\ref{eq:equal_t}. We re-write the CC Lagrangian (Eq.~\ref{eq:BCC_energy}), whilst using Eq.~\ref{eq:BCC_gf} and the effective interactions defined in Eqs ~\ref{eq:amp1},~\ref{eq:amp2} and so on, to get
\begin{gather}
    \begin{split}
        \mathcal{L}^{\BCC}_{c}[\tilde{G}] &= -\frac{i}{2}\sum_{pq}(F_{pq}-f_{pq})\tilde{G}_{qp} + \sum_{ai}\lambda^{i}_{a} F_{ai}\\
        &+\frac{1}{4}\sum_{abij}\lambda^{ij}_{ab}\chi_{ab,ij}+ \cdots
    \end{split}
\end{gather}
Taking the functional derivative of this expression with respect to the equal-time non-interacting Green's function gives 
\begin{gather}
    \begin{split}~\label{eq:exact_stat}
      \tilde{\Sigma}^{\infty}_{pq} &= \tilde{\Sigma}^{\infty (0)}_{pq} + \sum_{ai}\lambda^{i}_{a} \chi_{pa,qi}+\frac{1}{4}\sum_{abij}\lambda^{ij}_{ab}\chi_{pab,qij} + \cdots
    \end{split}
\end{gather}
where we have used the identities given in Eqs~\ref{eq:proof} and~\ref{eq:proof1} and so on for the higher-body interactions. The series in Eq.~\ref{eq:exact_stat} continues with the four-body interaction contracted with $\lambda^{ijk}_{abc}$ and so on. The first term of Eq.~\ref{eq:exact_stat} is given by
\begin{figure*}[ht]
    \centering
        \begin{gather*}
\begin{split}~\label{eq:exact_diags}
     \Tilde{\Sigma}^{\infty}_{pq} = \hspace{5mm}\begin{gathered}
\begin{fmfgraph*}(40,40)
    \fmfset{arrow_len}{3mm}
    \fmfleft{i1,i2,i3}
    \fmfright{o1,o2,o3}
    \fmf{fermion}{i1,i2}
    \fmf{fermion}{i2,i3}
    \fmf{dbl_dashes}{i2,o2}
    \fmfforce{(0.0w,0.h)}{i1}
    \fmfforce{(0.0w,0.5h)}{i2}
    \fmfforce{(0.0w,1.0h)}{i3}
    \fmfdot{i2}
    \fmfv{decor.shape=cross,decor.filled=full, decor.size=1.5thic}{o2}
\end{fmfgraph*}
\end{gathered}\hspace{5mm}+\hspace{7.5mm}
\begin{gathered}
    \begin{fmfgraph*}(40,40)
    \fmfcurved
    \fmfset{arrow_len}{3mm}
    \fmfleft{i1,i2}
    \fmflabel{}{i1}
    \fmflabel{}{i2}
    \fmfright{o1,o2}
    \fmflabel{}{o1}
    \fmflabel{}{o2}
    \fmf{dbl_wiggly}{i1,o1}
    \fmf{fermion,left=0.3,tension=0}{o1,o2}
    \fmf{fermion,left=0.3,tension=0}{o2,o1}
    \fmf{dbl_plain}{v1,o2}
    \fmf{dbl_plain}{o2,v2}
    \fmf{fermion}{v3,i1}
    \fmf{fermion}{i1,v4}
    \fmfforce{(1.05w,1.0h)}{v2}
    \fmfforce{(0.75w,1.0h)}{v1}
    \fmfforce{(0.0w,-0.5h)}{v3}
    \fmfforce{(0.0w,0.5h)}{v4}
    \fmfforce{(0.0w,0.0h)}{i1}
    \fmfdot{o1,i1}
\end{fmfgraph*}
\end{gathered}\hspace{5mm}+\hspace{7.5mm}
\begin{gathered}
    \begin{fmfgraph*}(60,40)
    \fmfcurved
    \fmfset{arrow_len}{3mm}
    \fmfleft{i1,i2}
    \fmflabel{}{i1}
    \fmflabel{}{i2}
    \fmfright{o1,o2}
    \fmflabel{}{o1}
    \fmflabel{}{o2}
    \fmf{dashes}{o1,v1}
    \fmf{dashes}{i1,v1}
    \fmf{fermion,left=0.3,tension=0}{o1,o2}
    \fmf{fermion,left=0.3,tension=0}{v1,v2}
    \fmf{fermion,left=0.3,tension=0}{v2,v1}
    \fmf{phantom}{v2,i2}
    \fmf{dbl_plain}{o2,v2}
    \fmf{fermion,left=0.3,tension=0}{o2,o1}
    \fmf{fermion}{v3,i1}
    \fmf{fermion}{i1,v4}
    \fmfdot{o1,i1,v1}
    \fmfforce{(0.0w,-0.5h)}{v3}
    \fmfforce{(0.0w,0.5h)}{v4}
    \fmfforce{(0.0w,0.0h)}{i1}
    \fmfforce{(0.5w,1.0h)}{v2}
    \fmfforce{(0.5w,0.0h)}{v1}
\end{fmfgraph*}
\end{gathered}\hspace{5mm}+\hspace{2.5mm}\cdots\\
\\
\end{split}
\end{gather*}
    \caption{Diagrammatic representation of the coupled-cluster self-energy defined in Eq.~\ref{eq:exact_stat}. The interactions are given in Eq.~\ref{eq:interactions}, with the dexcitation amplitudes given by the Lagrange multipliers from Eq.~\ref{eq:BCC_energy}.}
    \label{fig:full_se}
\end{figure*}
\begin{gather}
    \begin{split}
       \tilde{\Sigma}^{\infty(0)}_{pq} =i\frac{\delta E^{\BCC}_{c}}{\delta\tilde{G}_{qp}} ,
    \end{split}
\end{gather}
and using Eq.~\ref{eq:BCC_gf}, we find
\begin{gather}
    \begin{split}
        i\frac{\delta E^{\BCC}_{c}}{\delta\tilde{G}_{qp}} = \frac{1}{2}(F_{pq}-f_{pq}) -\frac{i}{2}\sum_{rs}\left(i\frac{\delta(F_{rs}-f_{rs})}{\delta\tilde{G}_{qp}}\right)\tilde{G}_{sr} \ .
    \end{split}
\end{gather}
Using Eqs~\ref{eq:fokk} and~\ref{eq:proof}, the second term yields
\begin{gather}
    \begin{split}~\label{eq:1st_kernel}
        i\frac{\delta(F_{rs}-f_{rs})}{\delta\tilde{G}_{qp}} = \chi_{pr,qs}-\braket{pr||qs} = \tilde{\Xi}^{c(0)}_{pr,qs} ,
    \end{split}
\end{gather}
and taking the trace of this quantity over the occupied states gives~\cite{shavitt2009many}
\begin{gather}
    \begin{split}
       -i\sum_{rs}\tilde{\Xi}^{c(0)}_{pr,qs}\tilde{G}_{sr}=\sum_{k}\tilde{\Xi}^{c(0)}_{pk,qk} = F_{pq}-f_{pq}\ . 
    \end{split}
\end{gather}
Combining both results together gives 
\begin{gather}
    \begin{split}~\label{eq:BCC_func}
        \tilde{\Sigma}^{\infty(0)}_{pq} = F_{pq}-f_{pq} \ .
    \end{split}
\end{gather}
The diagrammatic representation of Eq.~\ref{eq:exact_stat} is given in Fig.~\ref{fig:full_se}. The diagrammatic notation is as follows. The double lines indicate the dexcitation amplitudes $\{\lambda^{i}_{a},\lambda^{ij}_{ab},\cdots\}$ and so on, while the CC effective interactions are depicted as 
\begin{gather}
    \begin{split}~\label{eq:interactions}
        \begin{gathered}
            \begin{fmfgraph*}(40,40)
    \fmfcurved
    \fmfset{arrow_len}{3mm}
    \fmfleft{i1}
    \fmflabel{}{i1}
    \fmfright{o1}
    \fmf{dbl_wiggly}{i1,o1}
    \fmfdot{o1,i1}
\end{fmfgraph*}
        \end{gathered}\hspace{2.5mm}&= \chi_{pq,rs},\\
       \begin{gathered}
            \begin{fmfgraph*}(60,40)
    \fmfcurved
    \fmfset{arrow_len}{3mm}
    \fmfleft{i1}
    \fmflabel{}{i1}
    \fmfright{o1}
    \fmf{dashes}{i1,v1}
    \fmf{dashes}{v1,o1}
    \fmfdot{o1,i1,v1}
\end{fmfgraph*}
\end{gathered} \hspace{2.5mm}&= \chi_{pqr,stu} \ ,\\
\begin{gathered}
            \begin{fmfgraph*}(60,40)
    \fmfcurved
    \fmfset{arrow_len}{3mm}
    \fmfleft{i1}
    \fmflabel{}{i1}
    \fmfright{o1}
    \fmf{dbl_dashes}{i1,v1}
    \fmf{dbl_dashes}{v1,v2}
    \fmf{dbl_dashes}{v2,o1}
    \fmfdot{o1,i1,v1,v2}
\end{fmfgraph*}
\end{gathered} \hspace{2.5mm}&= \chi_{pqrw,stuv} \ ,
    \end{split}
\end{gather}
and so on, where we employ the antisymmetrized Hugenholtz-Shavitt-Bartlett diagram convention.~\cite{hirata2024nonconvergence} The first term, Eq.~\ref{eq:BCC_func}, is the generalization of Eq.~\ref{eq:se_ref} to the case of coupled-cluster theory. 

Using these results, we write the coupled-cluster quasiparticle equation from the resulting Dyson equation as
\begin{gather}
    \begin{split}~\label{eq:qp_ham}
        \tilde{F}_{pq} = f_{pq} + \tilde{\Sigma}^{\infty}_{pq}  \ .
    \end{split}
\end{gather}
To the best of our knowledge, Eq.~\ref{eq:exact_stat} contains higher-order contributions that have not been reported before. Focusing on the contribution to the coupled-cluster self-energy generated by the Brueckner correlation energy, $E^{\BCC}_c$ (via Eq.~\ref{eq:BCC_func}), the quasiparticle equation becomes 
\begin{gather}
    \begin{split}~\label{eq:ext_fock}
        \tilde{F}^{(0)}_{pq} = F_{pq} = f_{pq} + \tilde{\Sigma}^{\infty (0)}_{pq} \ .
    \end{split}
\end{gather}
From this expression, we immediately see that the self-energy for the occupied states is given by~\cite{scuseria1995connections,shavitt2009many} 
\begin{gather}
	\begin{split}\label{eq:se}
		 \Tilde{\Sigma}^{\infty(0)}_{ij} = \frac{1}{2} \sum_{kab} \braket{ik||ab}t^{ab}_{jk} \ ,
	\end{split}
\end{gather}
whilst the self-energy of the virtual states is~\cite{scuseria1995connections,shavitt2009many} 
\begin{equation}~\label{eq:virse}
	\Tilde{\Sigma}^{\infty(0)}_{ab} = -\frac{1}{2}\sum_{ijc} \braket{ij||bc}t^{ac}_{ij} \ .
\end{equation}
The self-energy for the occupied--virtual block in BCC theory vanishes identically as these are given by the $T_1$ amplitude equations, $\braket{\Phi^{a}_i|\Bar{H}|\Phi_0} = 0$.~\cite{scuseria1995connections} Therefore this definition of the self-energy decouples the IP and EA sectors. The expression, $\tilde{\Sigma}^{\infty}$, constitutes the most significant one-particle irreducible contribution to the coupled-cluster self-energy and are systematically improvable via the coupled-cluster hierarchy. The `zeroth-order' approximation, $\tilde{\Sigma}^{\infty(0)}_{ij}$, is exactly the correlation part of the generalized Fock operator introduced in Brueckner theory for the occupied states.~\cite{scuseria1995connections,scuseria2013particle,lange2018relation}

From Eq.~\ref{eq:se}, we obtain the exact ground state correlation energy by taking the trace of the self-energy
\begin{gather}
\begin{split}~\label{eq:GM_energy}
    E^\BCC_c &= -\frac{i}{2}\sum_{pq}\tilde{\Sigma}^{\infty(0)}_{pq}\tilde{G}_{qp}=\frac{1}{2}\sum_{i} \Tilde{\Sigma}^{\infty(0)}_{ii} \\
    &= \frac{1}{4} \sum_{ij,ab} \braket{ij||ab}t^{ab}_{ij} \ .
\end{split}
\end{gather} 
This is identical to the form of the RPA and $GW$-BSE correlation energy and originates from the Galitskii-Migdal formula (Eq.~\ref{eq:GM}) for a static self-energy.~\cite{scuseria2008ground,quintero2022connections,coveney2023regularized} From Eq.~\ref{eq:ext_fock}, the eigenvalues of the extended Fock operator automatically correspond to ionization energies and electron affinities, providing us with a correlated generalization of Koopman's theorem. The extended Fock operator corresponds to the normal-ordered similarity transformed Hamiltonian in the space of one hole/particle (1h)/(1p) Slater determinants 
\begin{equation}~\label{eq:ext_fok}
    F_{ij} = -\braket{\Phi_j|\Bar{H}_{N}|\Phi_i}  \hspace{2mm};\hspace{2mm} F_{ab}= \braket{\Phi^a|\Bar{H}_{N}|\Phi^b} ,
\end{equation}
\begin{figure*}[ht]
    \centering
    \begin{gather*}
        \begin{split}~\label{eq:bse_full}
            \tilde{\Xi}^{c}_{pq,rs} = \hspace{2.5mm}
\Bigg(\begin{gathered}
    \begin{fmfgraph*}(40,40)
    \fmfset{arrow_len}{3mm}
    \fmfleft{i1}
    \fmfright{o1}
    \fmf{dbl_wiggly}{o1,i1}
    \fmfdot{i1,o1}
\end{fmfgraph*}
\end{gathered}\hspace{2.5mm}-\hspace{2.5mm}\begin{gathered}
\begin{fmfgraph*}(40,40)
    \fmfset{arrow_len}{3mm}
    \fmfleft{i1}
    \fmfright{o1}
    \fmf{wiggly}{o1,i1}
    \fmfdot{i1,o1}
\end{fmfgraph*}
\end{gathered}\Bigg)\hspace{5mm}+\hspace{7.5mm}
\begin{gathered}
    \begin{fmfgraph*}(60,40)
    \fmfcurved
    \fmfset{arrow_len}{3mm}
    \fmfleft{i1,i2}
    \fmflabel{}{i1}
    \fmflabel{}{i2}
    \fmfright{o1,o2}
    \fmflabel{}{o1}
    \fmflabel{}{o2}
    \fmf{dashes}{o1,v1}
    \fmf{dashes}{i1,v1}
    \fmf{fermion,left=0.3,tension=0}{o1,o2}
    \fmf{phantom}{v2,i2}
    \fmf{fermion,left=0.3,tension=0}{o1,o2}
    \fmf{dbl_plain}{v3,o2}
    \fmf{dbl_plain}{o2,v4}
    \fmf{fermion,left=0.3,tension=0}{o2,o1}
    \fmf{fermion,left=0.3,tension=0}{o2,o1}
    \fmfdot{o1,i1,v1}
    \fmfforce{(1.05w,1.0h)}{v3}
    \fmfforce{(0.75w,1.0h)}{v4}
\end{fmfgraph*}
\end{gathered}\hspace{5mm}+\hspace{7.5mm}
\begin{gathered}
    \begin{fmfgraph*}(60,50)
    \fmfcurved
    \fmfset{arrow_len}{3mm}
    \fmfleft{i1,i2}
    \fmflabel{}{i1}
    \fmflabel{}{i2}
    \fmfright{o1,o2}
    \fmflabel{}{o1}
    \fmflabel{}{o2}
    \fmf{dbl_dashes}{v1,i1}
    \fmf{dbl_dashes}{o1,v1}
    \fmf{dbl_dashes}{v1,v3}
    \fmf{dbl_dashes}{v3,o1}
    \fmf{dbl_plain}{v4,o2}
    \fmf{fermion,left=0.3,tension=0}{o1,o2}
    \fmf{fermion,left=0.3,tension=0}{v3,v4}
    \fmf{fermion,left=0.3,tension=0}{v4,v3}
    \fmf{fermion,left=0.3,tension=0}{o2,o1}
    \fmfdot{o1,i1,v1,v3}
    \fmfforce{(0.0w,0.0h)}{i1}
    \fmfforce{(1.0w,0.0h)}{o1}
    \fmfforce{(0.25w,1h)}{v2}
    \fmfforce{(0.25w,0.0h)}{v1}
    \fmfforce{(0.625w,0.0h)}{v3}
    \fmfforce{(0.625w,1.0h)}{v4}
    \fmfforce{(0.0w,1.0h)}{i2}
    \fmfforce{(1.0w,1.0h)}{o2}
\end{fmfgraph*}
\end{gathered}\hspace{5mm}+\hspace{2.5mm} \cdots
        \end{split}
    \end{gather*}
    \caption{Diagrammatic representation of the coupled-cluster BSE kernel defined in Eq.~\ref{eq:exact_bse_stat}.}
    \label{fig:full_BSE}
\end{figure*}
where $\ket{\Phi_{i}} = a_i\ket{\Phi_0}$ and $\ket{\Phi^{a}} = a^\dag_a\ket{\Phi_0}$. These are exactly the elements appearing in exact IP/EA-EOM-CC theory. Importantly, many of the dynamical degrees of freedom of the self-energy have been re-expressed in terms of the static coupled-cluster amplitudes. In the supermatrix representation, the $GW$ and GF2 approximations couple the IP and EA sectors, implicitly containing orbital relaxation despite giving an inexact ground state energy and spectrum.~\cite{tolle2023exact,lange2018relation,bintrim2021full} Diagrammatically, the functional derivative of the Brueckner correlation  energy corresponds to cutting the hole/particle lines in the Goldstone diagram of the Brueckner correlation energy in Fig.~\ref{fig:SE} (see Appendix~\ref{app:gold}).

\section{Relationship between the coupled-cluster self-energy and Bethe-Salpeter kernel}~\label{sec:BSE}

Within the Green's function formalism, the Bethe-Salpeter equation describes two-particle correlation processes and gives access to neutral excitation energies. The two-particle Bethe-Salpeter interaction kernel is directly related to the self-energy and is given by the functional derivative~\cite{hedin1965new,rohlfing2000electron}
\begin{equation}
	\Xi^{c}(1,2;3,4) = i\frac{\delta\Sigma^{c}(1,3)}{\delta G(4,2)} \ .
\end{equation}
Importantly, the definition of the Bethe-Salpeter kernel is dependent on the choice of the self-energy approximation. Depending on the relative time-ordering of the field operators, the two-particle Green's function may describe particle-hole or particle-particle correlations.~\cite{scuseria2013particle,berkelbach2018communication} Focusing on the particle-hole propagator, we similarly define the analogous particle-hole interaction kernel from the CC self-energy defined in Eq.~\ref{eq:exact_stat} as
\begin{equation}
	\tilde{\Xi}^{c}_{ia,bj} = i\frac{\delta\Tilde{\Sigma}^{\infty}_{ib}}{\delta \Tilde{G}_{ja}}\ .
\end{equation}
This functional derivative gives the coupled-cluster two-particle interaction kernel from the Bethe-Salpeter equation (BSE) in the space of singly excited determinants. Identifying that the CC self-energy of Eq.~\ref{eq:exact_stat} also depends on the non-interacting Green's function allows us to take higher-order functional derivatives. The corresponding BSE kernel of the coupled-cluster self-energy defined in Eq.~\ref{eq:exact_stat} is given by  
\begin{gather}
    \begin{split}~\label{eq:exact_bse_stat}
    \tilde{\Xi}^{c}_{pq,rs} &=  i\frac{\delta \Tilde{\Sigma}^{\infty(0)}_{pr}}{\delta\tilde{G}_{sq}} + \sum_{ai} \lambda^{i}_{a}\chi_{pqa,rsi} + \frac{1}{4}\sum_{ijab}\lambda^{ij}_{ab}\chi_{pqab,rsij}\\
        &+\cdots 
    \end{split}
\end{gather}
with the series continuing with the five-body interaction contracted with $\lambda^{ijk}_{abc}$ and so on. The first term is given by Eq.~\ref{eq:1st_kernel} using the identities from Eqs.~\ref{eq:proof} and~\ref{eq:proof1}. The diagrammatic representation of this equation is depicted in Fig.~\ref{fig:full_BSE}, where the bare antisymmetrized Coulomb interaction is represented as
\begin{gather}
    \begin{split}
     \begin{gathered}
            \begin{fmfgraph*}(40,40)
    \fmfcurved
    \fmfset{arrow_len}{3mm}
    \fmfleft{i1}
    \fmflabel{}{i1}
    \fmfright{o1}
    \fmf{wiggly}{i1,o1}
    \fmfdot{o1,i1}
\end{fmfgraph*}
        \end{gathered}  \hspace{2.5mm}=\braket{pq||rs} \ .
    \end{split}
\end{gather}
The full kernel, including the variation of the reference self-energy, is given by 
\begin{gather}
    \begin{split}~\label{eq:exact_bse_stat_1}
    \tilde{\Xi}_{pq,rs} &= \chi_{pq,rs}  + \sum_{ai} \lambda^{i}_{a}\chi_{pqa,rsi} + \frac{1}{4}\sum_{ijab}\lambda^{ij}_{ab}\chi_{pqab,rsij}\\
        &+\cdots 
    \end{split}
\end{gather}
Neglecting the variation of higher-order effective interaction elements in Eq.~\ref{eq:exact_bse_stat}, we find that the `zeroth-order' expression for the CC-BSE kernel is given by 
\begin{gather}
    \begin{split}~\label{eq:kernel}
        \tilde{\Xi}^{(0)}_{ia,bj} = \braket{ia||bj}+\sum_{kc}\braket{ik||bc}t^{ca}_{jk} \ .
    \end{split}
\end{gather}
It should be noted that this kernel retains the fermionic antisymmetry of the many-body wavefunction via the doubles amplitudes. This term corresponds to the third diagram of Fig.~\ref{fig:SE}. From the Bethe-Salpeter equation, we obtain the CC-BSE Hamiltonian as
\begin{equation}
	\Bar{H}^{\BSE}_{ia,jb} = \tilde{F}_{ab}\delta_{ij} -\tilde{F}_{ij}\delta_{ab} +\Tilde{\Xi}_{ia,bj} \ .
\end{equation}
\begin{figure}[ht]
    \centering
    \includegraphics[width=83mm,height=60mm]{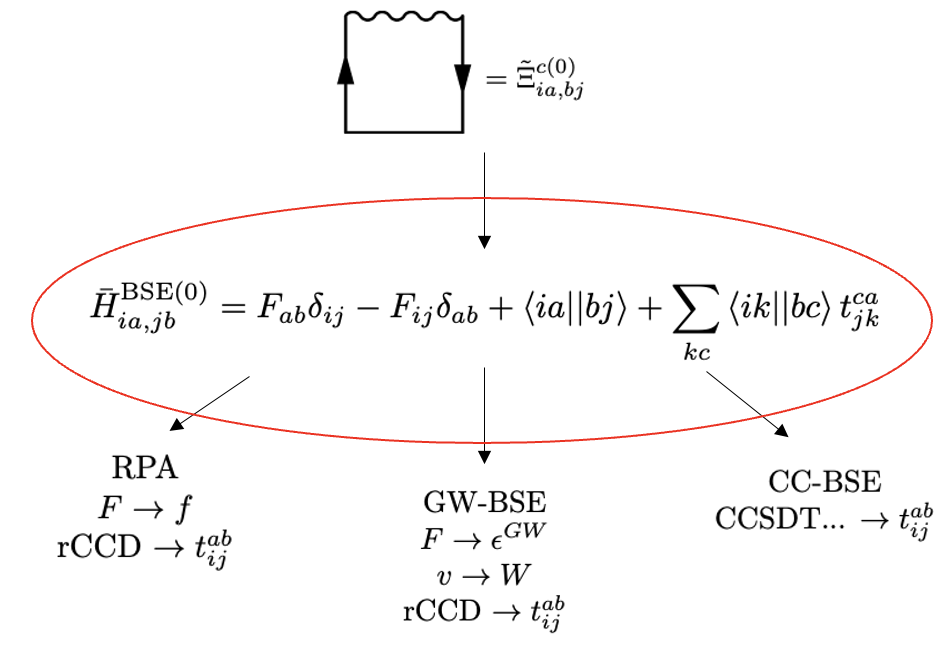}
    \caption{Relationship between the coupled-cluster self-energy, RPA, $GW$-BSE and CC-BSE.}
    \label{fig:rpa_rel}
\end{figure}
By replacing the effective interaction elements with their `zeroth-order' counterparts $\tilde{\Sigma}^{\infty(0)}_{ij}/\tilde{\Sigma}^{\infty(0)}_{ab}$ and $\tilde{\Xi}^{(0)}_{ia,bj}$ given in Eqs~\ref{eq:se}/~\ref{eq:virse} and Eq.~\ref{eq:kernel}, we may write the effective CC-BSE Hamiltonian in terms of the coupled-cluster doubles amplitudes as~\cite{berkelbach2018communication,scuseria2008ground,quintero2022connections} 
\begin{equation}~\label{eq:bse}
	\Bar{H}^{\BSE(0)}_{ia,jb} = F_{ab}\delta_{ij} -F_{ij}\delta_{ab} + \braket{ia||bj} + \sum_{kc}\braket{ik||bc}t^{ca}_{jk}
\end{equation}
whose eigenvalues give the exciton energies. Within this approximation, the CC-BSE Hamiltonian exactly corresponds to the normal-ordered similarity transformed Hamiltonian in the space of singly excited determinants~\cite{lange2018relation,berkelbach2018communication,rishi2020route}
\begin{gather}
	\begin{split}
		\Bar{H}^{\BSE(0)}_{ia,jb} &\equiv \braket{\Phi^a_j|\Bar{H}_N|\Phi^b_i} \ .
	\end{split}
\end{gather}
This is equivalent to the upper-left block from the excitation energy (EE)-EOM-CC treatment. Through the field theoretic identities derived above (Eqs~\ref{eq:kernel} and~\ref{eq:bse}), the relationship between IP/EA- and EE-EOM-CC theory is now explicitly clear.

The RPA and $GW$-BSE approximation are formulated in the space of singly excited determinants and the neutral excitation energies are obtained from the eigenvalue problem~\cite{ring2004nuclear,quintero2022connections,scuseria2008ground}    \begin{equation}\label{eq:rpa}
    \left(\begin{array}{cc}
        \mathbf{A} & \mathbf{B} \\
         -\mathbf{B}^* & -\mathbf{A}^*
    \end{array}\right)\left(\begin{array}{c}
          \mathbf{X} \\ 
          \mathbf{Y}
    \end{array}\right) = \left(\begin{array}{c}
          \mathbf{X} \\ 
          \mathbf{Y}
    \end{array}\right)\mathbf{\Omega} \ ,
\end{equation}
where 
\begin{subequations}
    \begin{gather}
        \begin{split}
            A_{ia,jb} = \Delta^{a}_{i}\delta_{ab}\delta_{ij}+ V_{ia,jb}
        \end{split}
    \end{gather}
    \begin{gather}
        \begin{split}
            B_{ia,jb} = V_{ij,ab} \ .
        \end{split}
    \end{gather}
\end{subequations}
Within the RPA: $\Delta^{a}_{i} = \epsilon_{a}-\epsilon_{i}$, $V_{ia,jb} = \braket{ia||bj}$ and $B_{ia,jb} = \braket{ij||ab}$. In the $GW$-BSE approach: $\Delta^{a}_{i} = \epsilon^{GW}_{a}-\epsilon^{GW}_{i}$, $V_{ia,jb} = \braket{ia|bj}-W_{ia,jb}$ and $B_{ia,jb} = \braket{ij|ab}-W_{ij,ba}$, where $W$ is the static screened interaction.~\cite{hybertsen1986electron,rohlfing2000electron} The corresponding eigenstates of Eq.~\ref{eq:rpa} are known as excitons, namely correlated electron-hole pairs. 

Scuseria, Henderson and Sorenson demonstrated that RPA reduces to a form of CCD where the doubles amplitudes are solved for by keeping only the so-called `ring' contractions (rCCD), and recently a similar structure for effective doubles amplitudes has been uncovered for $GW$-BSE.~\cite{scuseria2008ground,quintero2022connections} The relationship between RPA, $GW$-BSE and the doubles amplitudes is obtained by identifying that the doubles amplitudes are given by $t^{ab}_{ij} \equiv \mathbf{T} = \mathbf{YX}^{-1}$.~\cite{ring2004nuclear,scuseria2008ground} Using this relation, Eq.~\ref{eq:rpa} yields the following Ricciatti equation for the doubles amplitudes~\cite{scuseria2008ground}
\begin{equation}~\label{eq:rccd}
    \mathbf{B}^* + \mathbf{A^*T} + \mathbf{TA} + \mathbf{TBT} = 0 \ .
\end{equation}
These are exactly the rCCD amplitude equations first derived in Ref.~\citenum{scuseria2008ground}. The resulting RPA/$GW$-BSE equations can therefore be written as~\cite{berkelbach2018communication,quintero2022connections} 
\begin{equation}
    \mathbf{H}^{\RPA} = \mathbf{A} + \mathbf{BT} \ ,
\end{equation}
which, using the definition of the RPA matrix elements defined above, is written as
\begin{equation}~\label{eq:RPA_eqns}
    H_{ia,jb}^{\RPA} = \Delta^{a}_{i}\delta_{ij}\delta_{ab} + \braket{ia||bj} + \sum_{kc} \braket{ik||bc}t^{ca}_{jk} \ ,
\end{equation}
remembering that the doubles amplitudes are determined from the rCCD equations, which break fermionic antisymmetry (Eq.~\ref{eq:rccd}).

Here, we see the connection between our formalism and RPA in a simple and clear way. The structure of the RPA equations (Eq.~\ref{eq:RPA_eqns}) is exactly the same as the CC-BSE equations given by Eq.~\ref{eq:bse}. When higher-order excitation processes are neglected, as is the case for RPA, we see that the structure of our derived CC-BSE effective Hamiltonian $\Bar{H}^{\BSE}$, is identical to the RPA/$GW$-BSE description. Our general formalism reduces to the RPA eigenvalue problem when the effects of the self-energy on the single-particle states are neglected and the doubles amplitudes are solved via the rCCD equations.~\cite{rishi2020route,berkelbach2018communication,scuseria2013particle} Likewise, the $GW$-BSE approximation can be obtained by using the $GW$ eigenvalues for the valence and conduction bands and the screened instead of bare interaction in the rCCD amplitude equations.~\cite{scott2023moment,bintrim2021full,bintrim2022full} Therefore, in RPA and $GW$-BSE, the rCCD amplitudes do not include additional contributions from the full CC doubles amplitude equations such as ladder diagrams, triple excitations and so on (Fig.~\ref{fig:rpa_rel}). Interestingly, our formulation  provides a systematic way to include three body interactions via perturbative triples amplitudes (or better). As a result, increasing the interactions contained in the doubles amplitude equations can be likened to an iterative approach for including vertex corrections to the self-energy.~\cite{schindlmayr1998systematic,molinari2005hedin,ness2011g,lani2012approximations,kutepov2017self,harkov2021parametrizations,kutepov2022full,mejuto2022self} This correspondence can be identified from the structure of the doubles amplitude equations which include higher-order correlation effects depending on the truncation of the cluster operator, $T$. Inclusion of higher-order excitations in the cluster operator and iterative solution of the resulting doubles amplitude equations systematically includes higher-order correlation effects analogous to the vertex function appearing in Hedin's equations. 
The equations derived here demonstrate the equivalence between the BSE, ground state CC and EOM-CC theory. 

The exact ground state correlation energy is obtained by taking the trace over the BSE kernel
\begin{gather}
	\begin{split}
		E^{\BCC}_{c} = \frac{1}{4}\tr\mathbf{\tilde{\Xi}}^{c(0)} = \frac{1}{4}\sum_{ijab}\braket{ij||ab}t^{ab}_{ij}\ . 
	\end{split}
\end{gather}
This represents a formally exact extension of the elegant proof of the relationship between the RPA and CCD ground state correlation energies to yield the exact ground state correlation energy.~\cite{scuseria2008ground} 

To conclude this section, we also define the particle-particle and hole-hole two-particle interaction kernels analogously to give 
\begin{subequations}
    \begin{equation}
        \tilde{\Xi}^{(0)}_{ab,cd} = \braket{ab||cd} + \sum_{i<j}\braket{ij||cd}t^{ab}_{ij}
    \end{equation}
    and 
     \begin{equation}
        \tilde{\Xi}^{(0)}_{ij,kl} = \braket{ij||kl} + \sum_{a<b}\braket{ij||ab}t^{ab}_{kl} ,
    \end{equation}
\end{subequations}
respectively.~\cite{berkelbach2018communication} Their corresponding CC-BSEs are given by the matrix elements of the similarity transformed Hamiltonian in the basis of 2p or 2h determinants. By continuation of the series of functional derivatives, we also obtain the three-body interaction kernel
\begin{equation}
    \tilde{\chi}_{pqr,stu} =  i\frac{\delta\tilde{\Xi}_{pq,st}}{\delta\Tilde{G}_{ur}} \ . 
\end{equation}
Taking this functional derivative gives rise to the diagrammatic series 
\begin{gather}
    \begin{split}~\label{eq:exact_3bod_stat}
    \tilde{\chi}_{pqr,stu} &=  \chi_{pqr,stu} + \sum_{ai} \lambda^{i}_{a}\chi_{pqra,stui} \\
    &+ \frac{1}{4}\sum_{ijab}\lambda^{ij}_{ab}\chi_{pqrab,stuij}+\cdots 
    \end{split}
\end{gather}
which is depicted in Eq.~\ref{eq:3-bod} of Appendix~\ref{app:dyn_ccse}.
Restricting the components of the three-body kernel to the two-hole, one-particle sector (2h1p) gives
\begin{equation}~\label{eq:2h1p_kernel}
    \tilde{\chi}_{ija,kbl} =  i\frac{\delta\tilde{\Xi}_{ja,bl}}{\delta\Tilde{G}_{ki}} \ ,
\end{equation}
which corresponds to the kernel of the 2h1p Green's function.~\cite{riva2023multichannel} Neglecting the variation of the higher-order effective interaction elements in Eq.~\ref{eq:2h1p_kernel}, at `zeroth-order' the 2h1p interaction kernel is 
\begin{equation}
    \tilde{\chi}^{(0)}_{ija,kbl} =\chi_{ija,kbl} = \sum_{c} \braket{ij||bc}t^{ca}_{lk}  \ .
\end{equation}
Diagrammatically, this term is given by 
\begin{gather}
    \begin{split}
        \tilde{\chi}^{(0)}_{ija,kbl} =\hspace{5mm}
        \begin{gathered}
            \begin{fmfgraph*}(40,40)
\fmfset{arrow_len}{3mm}
    \fmfleft{i1,i2}
    \fmfright{o1,o2}
    \fmf{plain}{i1,o1}
    \fmf{fermion}{i1,v1}
    \fmf{fermion}{v2,i1}
    \fmf{fermion}{v3,o1}
    \fmf{fermion}{i2,v4}
    \fmf{fermion}{v5,i2}
    \fmf{wiggly}{i2,o2}
    \fmf{fermion}{o2,v6}
    \fmf{fermion}{o1,o2}
    \fmfdot{i2,o2}
    \fmfforce{(1.75w,0.0h)}{i1}
    \fmfforce{(1.65w,0.5h)}{v2}
    \fmfforce{(1.95w,0.5h)}{v1}
    \fmfforce{(1.25w,0.5h)}{v3}
    \fmfforce{(0.25w,0.65h)}{v4}
    \fmfforce{(-0.1w,0.65h)}{v5}
    \fmfforce{(0.7w,0.65h)}{v6}
\end{fmfgraph*}
        \end{gathered} 
    \end{split}
\end{gather}
and corresponds to the third derivative of the Brueckner correlation energy (see Appendix~\ref{app:gold}). These kernels will be useful in deriving the relationship between the coupled-cluster self-energy and IP/EA-EOM-CCSD in Section~\ref{sec:rel_eomcc}.

\section{Relationship to IP/EA-EOM-CC theory}~\label{sec:rel_eomcc}

To uncover the relationship between the coupled-cluster self-energy and IP/EA-EOM-CC theory, we first introduce the spectral form of the coupled-cluster self-energy and the coupled-cluster Dyson supermatrix, whose eigenvalues give the exact charged excitation energies. We then proceed to show how replacing the interaction kernels derived in this work with their perturbative counterparts results in the equivalence between the IP/EA-EOM-CCSD eigenvalue problem and the coupled-cluster self-energy. We emphasise here that both the untruncated coupled-cluster self-energy and IP/EA-EOM-CC theories are formally exact approaches, but with differing mathematical structures. 

\subsection{The dynamical coupled-cluster self-energy}

The irreducible coupled-cluster self-energy is composed of all one-particle irreducible (1PI) Feynman diagrams constructed from effective interactions generated by the similarity transformed Hamiltonian. By definition, the diagrammatic method (the set of all one-particle irreducible diagrams) yields an eigenvalue problem that is formally exact for the charged excitation spectrum of any general underlying many-body Hamiltonian. As the similarity transformed Hamiltonian is constructed by similarity transformation of the electronic structure Hamiltonian, their spectra are identical. The spectral form of the non-hermitian dynamical coupled-cluster self-energy is given by
\begin{gather}
    \begin{split}~\label{eq:spec_se_cc}
        \Tilde{\Sigma}_{pq}(\omega) &= \Tilde{\Sigma}^{\infty}_{pq} + \sum_{JJ'} \tilde{U}_{pJ}\left[(\omega+i\eta)\mathbbm{1}-(\mathbf{\bar{K}}^{>}+\mathbf{\bar{C}}^{>})\right]^{-1}_{JJ'}\bar{U}_{J'q}\\
        &+ \sum_{AA'} \bar{V}_{pA}\left[(\omega-i\eta)\mathbbm{1}-(\mathbf{\bar{K}}^{<}+\mathbf{\bar{C}}^{<})\right]^{-1}_{AA'}\tilde{V}_{A'q} \ .
    \end{split}
\end{gather}
Eq.~\ref{eq:spec_se_cc} is a formally exact representation of the coupled-cluster self-energy, containing the set of all possible 1PI diagrams. The first term of Eq.~\ref{eq:spec_se_cc} is the exact static component derived in Section~\ref{sec:sim_mbgf} (see Eq.~\ref{eq:exact_stat}). The second and third terms represent dynamical forward- and backward-time contributions to the coupled-cluster self-energy that couple the initial and final single-particle states (the external legs of a self-energy diagram) to multi-particle-hole excitations. Using Eq.~\ref{eq:spec_se_cc}, the frequency-dependent quasiparticle equation (Eq.~\ref{eq:qp_Eq.}) is equivalent to diagonalization of the coupled-cluster Dyson supermatrix
\begin{gather}
    \begin{split}~\label{eq:cc_dyson}
        \mathbf{\tilde{D}}^{\CC} = \left(\begin{array}{ccc}
            f_{pq}+\tilde{\Sigma}^{\infty}_{pq} & \tilde{U}_{pJ'} & \bar{V}_{pA'} \\
            \bar{U}_{Jq} & (\mathbf{\bar{K}}^{>}_{JJ'}+\mathbf{\bar{C}}^{>}_{JJ'}) & \mathbf{0} \\
             \tilde{V}_{Aq} & \mathbf{0} & (\mathbf{\bar{K}}^{<}_{AA'}+\mathbf{\bar{C}}^{<}_{AA'}) 
        \end{array}\right) .
    \end{split}
\end{gather}
The indices $JJ'/AA'$ label forward-time/backward-time ISCs. Here, the coupling matrices, $\bar{U}_{pJ}/\tilde{U}_{Jp}$ and $\bar{V}_{pA}/\tilde{V}_{Ap}$, mediate the coupling of the initial and final single-particle states to the ISCs induced by the similarity transformed Hamiltonian, $\bar{H}$. The interaction matrices, $(\mathbf{\bar{K}}^{>}_{JJ'}+\mathbf{\bar{C}}^{>}_{JJ'})$ and $(\mathbf{\bar{K}}^{<}_{AA'}+\mathbf{\bar{C}}^{<}_{AA'})$, represent the interactions between the different ISCs mediated by $\bar{H}$. We will subsequently demonstrate that the `interaction matrices' appearing in IP/EA-EOM-CCSD can be related to the 2h1p/2p1h matrix elements of the similarity transformed Hamiltonian (see Eq.~\ref{eq:eom_connect}). Importantly, for the non-hermitian coupled-cluster self-energy, the coupling matrices are not simply related to each other via hermitian conjugation. This is to be contrasted with the coupling matrices of the electronic Dyson supermatrix defined in Eq.~\ref{eq:el_dys}.
Similarly, the upper left block of $\mathbf{\tilde{D}}^{\CC}$ is defined over the complete set of occupied and virtual single-particle spin-orbitals. Diagonalization of the exact coupled-cluster Dyson supermatrix (Eq.~\ref{eq:cc_dyson}) yields the complete set of exact ionization potentials and electron affinities.

\subsection{Connection to IP/EA-EOM-CCSD eigenvalue problem}
Now, we identify the relationship between the interaction kernels derived in this work and the IP-EOM-CCSD eigenvalue problem in order to expose the relationship between IP-EOM-CC theory and the coupled-cluster self-energy for a particular case.

To expose the connection to IP-EOM-CCSD theory, we truncate the ISCs to be of 2p1h/2h1p character only and restrict our analysis to the occupied-occupied block of the self-energy (see Appendix~\ref{app:dyn_ccse} for the CC self-energy defined over the combined space of occupied and virtual orbitals). In this case, the coupling matrices that connect the initial and final single-particle states to the ISCs are given by variation of the quasiparticle Hamiltonian defined in Eq.~\ref{eq:qp_ham}. Focusing on the coupling matrices that connect the ISCs to initial and final single-particle states in the occupied-occupied subspace, we have 
\begin{subequations}
    \begin{gather}
        \begin{split}
            \bar{V}_{i,kla} &= i\frac{\delta \tilde{F}_{ik}}{\delta\tilde{G}_{la}} = \tilde{\Xi}_{ia,kl}
        \end{split}
    \end{gather}
    \begin{gather}
        \begin{split}
            \tilde{V}_{mnb,j} &= i\frac{\delta \tilde{F}_{mj}}{\delta\tilde{G}_{bn}} = \tilde{\Xi}_{mn,jb} 
        \end{split}
    \end{gather}
    \begin{gather}
    \begin{split}
        \tilde{U}_{i,kab} = i\frac{\delta \tilde{F}_{ia}}{\delta\tilde{G}_{bk}} = \tilde{\Xi}_{ik,ab}
    \end{split}
\end{gather}
\begin{gather}
    \begin{split}~\label{eq:vanish}
        \bar{U}_{lcd,j} = i\frac{\delta \tilde{F}_{cj}}{\delta\tilde{G}_{ld}} = \tilde{\Xi}_{cd,jl} \ ,
    \end{split}
\end{gather}
\end{subequations}
where the quantities $\tilde{\Xi}_{pq,rs}$ are defined in Eq.~\ref{eq:exact_bse_stat_1}. Importantly, the effective interaction 
\begin{gather}
    \begin{split}
        \tilde{\Xi}_{ab,ij} &=  \chi_{ab,ij} + \sum_{ck} \lambda^{k}_{c}\chi_{abc,ijk} + \frac{1}{4}\sum_{klcd}\lambda^{kl}_{cd}\chi_{abcd,ijkl}\\
        &+ \cdots \\
        &= 0
    \end{split}
\end{gather}
vanishes as a result of the amplitude equations, $\braket{\Phi_{\mu}|\bar{H}_N|\Phi_0} = 0$, represented by $\{\chi_{ab,ij},\chi_{abc,ijk},\chi_{abcd,ijkl},\cdots\}$. Therefore, upon downfolding the coupled-cluster Dyson supermatrix back into the single-particle spin-orbital subspace, the forward-time contribution to the occupied-occupied block of the coupled-cluster self-energy, arising due to the coupling to the 2p1h configurations (Eq.~\ref{eq:vanish}), vanishes as: $\bar{U}_{lcd,j}=\tilde{\Xi}_{cd,jl} = 0 $. Using the perturbative expression $\tilde{\Sigma}^{\infty}_{pq}\approx\tilde{\Sigma}^{\infty(0)}_{pq}$, the coupling matrices reduce to
\begin{subequations}
    \begin{gather}
        \begin{split}~\label{eq:first_order}
            \bar{V}^{(0)}_{i,kla} &= i\frac{\delta F_{ik}}{\delta\tilde{G}_{la}} = \braket{ia||kl}+i\frac{\delta\tilde{\Sigma}^{\infty(0)}_{ik}}{\delta\tilde{G}_{la}} = \chi_{ia,kl}
        \end{split}
    \end{gather}
    \begin{gather}
        \begin{split}~\label{eq:first_order1}
            \tilde{V}^{(0)}_{mnb,j} &= i\frac{\delta F_{mj}}{\delta\tilde{G}_{bn}} = \braket{mn||jb}+i\frac{\delta\tilde{\Sigma}^{\infty(0)}_{mj}}{\delta\tilde{G}_{bn}} = \chi_{mn,jb} \ ,
        \end{split}
    \end{gather}
\end{subequations}
where we have used Eqs.~\ref{eq:proof} and~\ref{eq:proof1}. 

To find the matrices that represent the interactions between the 2h1p ISCs, we follow the approach outlined in Ref.~\citenum{riva2023multichannel} and allow the particles and holes to interact through the CC self-energy (Eq.~\ref{eq:exact_stat}), pairwise via the CC-BSE kernel (Eq.~\ref{eq:exact_bse_stat_1}) and through the explicit three-body 2h1p interaction kernel (Eq.~\ref{eq:2h1p_kernel}), while correctly accounting for fermionic exchange symmetry, to give
\begin{gather}
    \begin{split}~\label{eq:backward}
        (\mathbf{\bar{K}}^{<}_{ija,klb}+\mathbf{\bar{C}}^{<}_{ija,klb}) &=   \tilde{F}_{ik}\delta_{ab}\delta_{jl}+\tilde{F}_{jl}\delta_{ab}\delta_{ik}-\tilde{F}_{ba}\delta_{ik}\delta_{jl}\\
    &-\tilde{F}_{il}\delta_{ab}\delta_{jk}-\tilde{F}_{jk}\delta_{ab}\delta_{il}+\tilde{F}_{ba}\delta_{il}\delta_{jk}\\
&+\tilde{\Xi}_{ib,al}\delta_{jk}+\tilde{\Xi}_{jb,ak}\delta_{il}- \tilde{\Xi}_{ij,kl}\delta_{ab} \\
    &-\tilde{\Xi}_{ib,ak}\delta_{jl}- \tilde{\Xi}_{jb,al}\delta_{ik} + \tilde{\chi}_{ijb,kal} \ .
    \end{split}
\end{gather}
The interaction matrices between the 2p1h ISCs may be defined analogously (see Appendix~\ref{app:dyn_ccse}). Therefore, focusing on the occupied-occupied contribution to the coupled-cluster Dyson supermatrix, while restricting the coupling to the space of 2h1p/2p1h ISCs, we have 
\begin{gather}
    \begin{split}~\label{eq:dys_occ}
        \mathbf{\tilde{D}}^{\CC}_{\occ} = \left(\begin{array}{cc}
            f_{ij}+\tilde{\Sigma}^{\infty}_{ij}  & \tilde{\Xi}_{ia,kl} \\
             \tilde{\Xi}_{mn,jb} & \mathbf{\bar{K}}^{<}_{mnb,kla}+\mathbf{\bar{C}}^{<}_{mnb,kla}
        \end{array}\right) ,
    \end{split}
\end{gather}
where the forward time contribution to the occupied-occupied block of the self-energy vanishes due to Eq.~\ref{eq:vanish}. Downfolding the occupied-occupied contribution to the coupled-cluster Dyson supermatrix (Eq.~\ref{eq:dys_occ}), we generate the CC self-energy (restricted to the space of 2p1h/2h1p excitations) for the occupied-occupied block as
\begin{gather}
    \begin{split}~\label{eq:cc_se_full}
        &\tilde{\Sigma}^{2\p1\h/2\h1\p}_{ij}(\omega) = \tilde{\Sigma}^{\infty}_{ij} + \frac{1}{4} \sum_{\substack{kla\\mnb}}\tilde{\Xi}_{ia,kl}\\
        &\times\left[(\omega-i\eta)\mathbbm{1}-(\mathbf{\bar{K}}^{<}+\mathbf{\bar{C}}^{<})\right]^{-1}_{kla,mnb}\tilde{\Xi}_{mn,jb} \ .
    \end{split}
\end{gather}
The 1PI diagrams contained in Eq.~\ref{eq:cc_se_full} are presented in Appendix~\ref{app:dyn_ccse}. The connection between the coupled-cluster self-energy and the IP-EOM-CCSD eigenvalue problem emerges first when the ISC interaction matrices are truncated to `zeroth-order' as
\begin{gather}
    \begin{split}
        (\mathbf{\bar{K}}^{<(0)}_{ija,klb}+\mathbf{\bar{C}}^{<(0)}_{ija,klb}) &=   F_{ik}\delta_{ab}\delta_{jl}+F_{jl}\delta_{ab}\delta_{ik}-F_{ba}\delta_{ik}\delta_{jl}\\
    &-F_{il}\delta_{ab}\delta_{jk}-F_{jk}\delta_{ab}\delta_{il}+F_{ba}\delta_{il}\delta_{jk}\\
&+\tilde{\Xi}^{(0)}_{ib,al}\delta_{jk}+\tilde{\Xi}^{(0)}_{jb,ak}\delta_{il}- \tilde{\Xi}^{(0)}_{ij,kl}\delta_{ab} \\
    &-\tilde{\Xi}^{(0)}_{ib,ak}\delta_{jl}- \tilde{\Xi}^{(0)}_{jb,al}\delta_{ik} + \tilde{\chi}^{(0)}_{ijb,kal} \ .
    \end{split}
\end{gather}
The sum of all these terms is exactly equal to the interaction between the 2h1p ISCs mediated by the similarity transformed Hamiltonian as
\begin{gather}
    \begin{split}~\label{eq:eom_connect}
        &(\mathbf{\bar{K}}^{<(0)}_{ija,klb}+\mathbf{\bar{C}}^{<(0)}_{ija,klb})=-\braket{\Phi^{b}_{kl}|\bar{H}_N|\Phi^{a}_{ij}}  \ .
    \end{split}
\end{gather} 
Then, by restricting the static contribution $\tilde{\Sigma}^{\infty}_{ij}$ to $\tilde{\Sigma}^{\infty(0)}_{ij}$ as defined in Eq.~\ref{eq:se} (neglecting the full variation of the CC Lagrangian) as well as the coupling matrices to their  `zeroth-order' counterparts defined in Eqs~\ref{eq:first_order} and~\ref{eq:first_order1}, the coupled-cluster Dyson supermatrix becomes 
\begin{gather}
    \begin{split}~\label{eq:approx_dys}
        \mathbf{\tilde{D}}^{\CC(0)}_{\occ} = \left(\begin{array}{cc}
            f_{ij}+\tilde{\Sigma}^{\infty(0)}_{ij}  & \chi_{ia,kl} \\
             \chi_{mn,jb} & -\braket{\Phi^{b}_{mn}|\bar{H}_N|\Phi^{a}_{kl}}
        \end{array}\right) \ .
    \end{split}
\end{gather}
From Eq.~\ref{eq:ext_fok}, the upper left block of Eq.~\ref{eq:approx_dys} corresponds to the extended Fock operator and identifying that $\chi_{ia,kl}=\braket{\Phi^{a}_{kl}|\Bar{H}_N|\Phi_{i}}$ and $\chi_{mn,jb}=\braket{\Phi_{j}|\Bar{H}_N|\Phi^{b}_{mn}}$,~\cite{shavitt2009many} we have 
\begin{gather}
    \begin{split}~\label{eq:approx_dys_eom}
        \mathbf{\tilde{D}}^{\CC(0)}_{\occ} = \left(\begin{array}{cc}
            -\braket{\Phi_{j}|\Bar{H}_N|\Phi_{i}}  & \braket{\Phi^{a}_{kl}|\bar{H}_N|\Phi_{i}} \\
             \braket{\Phi_{j}|\bar{H}_N|\Phi^{b}_{mn}} & -\braket{\Phi^{b}_{mn}|\bar{H}_N|\Phi^{a}_{kl}}
        \end{array}\right) \ .
    \end{split}
\end{gather}
Downfolding Eq.~\ref{eq:approx_dys_eom} results in the expression for the coupled-cluster self-energy in the occupied-occupied subspace as
\begin{gather}
    \begin{split}~\label{eq:occ_cc_se}
        &\tilde{\Sigma}^{2\p1\h/2\h1\p(0)}_{ij}(\omega) = \tilde{\Sigma}^{\infty(0)}_{ij} + \frac{1}{4} \sum_{\substack{kla\\mnb}}\chi_{ia,kl}\\
        &\times\left[(\omega-i\eta)\mathbbm{1}-(\mathbf{\bar{K}}^{<(0)}+\mathbf{\bar{C}}^{<(0)})\right]^{-1}_{kla,mnb}\chi_{mn,jb} \ . 
    \end{split}
\end{gather}

As outlined in Section~\ref{sec:eom_cc}, the IP-EOM-CC eigenvalue problem is defined by the non-hermitian supermatrix of Eq.~\ref{eq:eom_bcc}. Truncating $R^{\IP}_{\nu}$, as defined in Eq.~\ref{eq:eom}, to the space of 2h1p excitations (2p1h excitations in the case of EA-EOM-CC theory), we have the standard IP-EOM-CCSD supermatrix
\begin{gather}
    \begin{split}~\label{eq:eom_ccsd}
       -\mathbf{\bar{H}}^{\text{IP-EOM-CCSD}} = \left(\begin{array}{cc}
           \braket{\Phi_{i}|\bar{H}_N|\Phi_{j}}  & \braket{\Phi_{i}|\bar{H}_N|\Phi^{a}_{kl}}\\
            \braket{\Phi^{b}_{mn}|\bar{H}_N|\Phi_{j}} & \braket{\Phi^{b}_{mn}|\bar{H}_N|\Phi^{a}_{kl}} \\
        \end{array}\right) .
    \end{split}
\end{gather}
Downfolding this eigenvalue problem into the occupied-occupied subspace, using a Løwdin partitioning,~\cite{lowdin1982partitioning} we obtain the frequency-dependent effective Hamiltonian
\begin{gather}
\begin{split}
   &\bar{H}^{\text{EOM-CCSD}}_{ij}(\omega) = -\braket{\Phi_{i}|\bar{H}_N|\Phi_{j}}\\
        &+\frac{1}{4}\sum_{\substack{kla\\mnb}}\braket{\Phi_{i}|\Bar{H}_N|\Phi^{b}_{mn}}\left(\omega\mathbbm{1}-(\mathbf{\bar{K}}^{<(0)}+\mathbf{\bar{C}}^{<(0)})\right)^{-1}_{mnb,kla}\\
        &\hspace{15mm}\times\braket{\Phi^{a}_{kl}|\Bar{H}_N|\Phi_{j}} , 
\end{split} 
\end{gather}
where we have used Eq.~\ref{eq:eom_connect} for the interaction matrices. The negative signs appearing in the `off-diagonal' coupling terms, $\braket{\Phi_{i}|\Bar{H}_N|\Phi^{b}_{mn}}$ and $\braket{\Phi^{a}_{kl}|\Bar{H}_N|\Phi_{j}}$, of the supermatrix $\mathbf{\bar{H}}^{\text{IP-EOM-CCSD}}$ vanish as a result of the downfolding procedure. From Eq.~\ref{eq:ext_fok}, the first term evaluates to $-\braket{\Phi_{i}|\bar{H}_N|\Phi_{j}}=f_{ji}+\tilde{\Sigma}^{\infty(0)}_{ji}$ and so the IP-EOM-CCSD effective Hamiltonian becomes 
\begin{gather}
    \begin{split}~\label{eq:ip_eom_se}
        &\bar{H}^{\text{EOM-CCSD}}_{ij}(\omega) =  f_{ji}+\tilde{\Sigma}^{\infty(0)}_{ji} + \frac{1}{4} \sum_{\substack{kla\\mnb}}\chi_{ja,kl}\\
        &\times\left(\omega\mathbbm{1}-(\mathbf{\bar{K}}^{<(0)}+\mathbf{\bar{C}}^{<(0)})\right)^{-1}_{kla,mnb}\chi_{mn,ib} ,
    \end{split}
\end{gather}
where we have again used $\chi_{ja,kl}=\braket{\Phi^{a}_{kl}|\Bar{H}_N|\Phi_{j}}$ and $\chi_{mn,ib}=\braket{\Phi_{i}|\Bar{H}_N|\Phi^{b}_{mn}}$.~\cite{shavitt2009many} By taking the transpose of Eq.~\ref{eq:ip_eom_se} and identifying the effective EOM-CCSD `non-Dyson self-energy' as $\bar{\Sigma}^{\text{EOM-CCSD}}_{ij}(\omega) = \bar{H}^{\text{EOM-CCSD}}_{ji}(\omega)-f_{ij}$, we have
\begin{gather}
    \begin{split}~\label{eq:ip_eom_eff}
        &\bar{\Sigma}^{\text{EOM-CCSD}}_{ij}(\omega) =  \tilde{\Sigma}^{\infty(0)}_{ij} + \frac{1}{4} \sum_{\substack{kla\\mnb}}\chi_{ia,kl}\\
        &\times\left(\omega\mathbbm{1}-(\mathbf{\bar{K}}^{<(0)}+\mathbf{\bar{C}}^{<(0)})\right)^{-1}_{kla,mnb}\chi_{mn,jb} .
    \end{split}
\end{gather}
 By comparing Eqs~\ref{eq:occ_cc_se} and~\ref{eq:ip_eom_eff}, we see the equivalence between the approximated coupled-cluster self-energy in the occupied-occupied subspace and the IP-EOM-CCSD `non-Dyson self-energy'. The analysis presented here equally applies to the EA-EOM-CCSD eigenvalue problem and the virtual-virtual block of the coupled-cluster self-energy, thereby demonstrating how our formalism reduces to IP/EA-EOM-CCSD theory. It is important to note that downfolding the IP/EA-EOM-CC eigenvalue problem into the occupied-occupied/virtual-virtual subspace results in an \emph{effective} `non-Dyson self-energy' that is only defined over the occupied-occupied/virtual-virtual spin-orbitals.~\cite{lange2018relation} This is in contrast to the coupled-cluster self-energy which contains components over the larger combined space of occupied and virtual orbitals as it is derived from Feynman diagrammatic perturbation theory. However, the diagrammatic coupled-cluster self-energy and algebraic-wavefunction based IP/EA-EOM-CC methods can both be made formally exact despite their different mathematical formulations.

\section{Conclusions and Outlook}~\label{sec:conc}

In summary, we have introduced a natural definition of the coupled-cluster self-energy that unifies elements of coupled-cluster theory with the Green's function formalism. By appealing to the coupled-cluster similarity transformed Hamiltonian, we present a closed form expression for the self-energy that is systematically improvable via the coupled-cluster amplitude equations. We have discovered that the EOM-CC treatment naturally emerges as a consequence of our formalism, providing us with a unified description of RPA, $GW$-BSE and coupled-cluster theory for ground state and excitation energies. Finally, we have explicitly demonstrated how the IP/EA-EOM-CC eigenvalue problem is related to the diagrammatic coupled-cluster self-energy. The connections uncovered here represent a major step towards a unified description of the seemingly disconnected approaches of coupled-cluster theory and the Green's function formalism. We hope that our findings will stimulate renewed work towards an improved formulation of Green's function theory for correlated excited state phenomena.

\section*{Acknowledgments}

The authors are grateful to the anonymous referee for providing insightful comments during the review process. 

\appendix

\section{Effective interactions of the similarity transformed Hamiltonian and many-body Green's functions}~\label{app:1}

Here, we present the proof of Eq.~\ref{eq:proof} in the main text. The proof of Eq.~\ref{eq:proof1} is analogous. The `non-interacting' 2ph Green's function, $G^{2\p\h,(0)}$, is related to the `non-interacting' 4-point Green's function via 
    \begin{gather}
    \begin{split}
        &iG^{4\pt,(0)}_{pq,rs}(t_1,t_2;t_3,t_4) = \\
        &\braket{\Phi_0|\mathcal{T}\left\{a_{q}(t_2) a_p(t_1) a^\dag_{r}(t_3)a^\dag_{s}(t_4)\right\}|\Phi_0}
    \end{split}
\end{gather}
with the property
    \begin{gather}
    \begin{split}
        G^{4\pt,(0)}_{pq,rs}(t^+,t;t',t'^{+}) = G^{2\p\h (0)}_{pq,rs}(t-t') \ .
    \end{split}
\end{gather}
The equal time `non-interacting' 2ph Green's function function yields the `non-interacting' two-particle reduced density matrix via
\begin{equation}
    \Gamma^{(0)}_{pq,rs} = iG^{2\p\h,(0)}_{pq,rs}(t-t^{+}) = \braket{\Phi_0|a^\dag_{r}a^\dag_s a_q a_p|\Phi_0} \ .
\end{equation}
As a result, the non-interacting 2ph Green's function is related to the equal-time `non-interacting' single-particle Green's function via 
\begin{equation}
    iG^{2\p\h,(0)}_{pq,rs}(t-t^{+}) = \tilde{G}_{qr}\tilde{G}_{ps}-\tilde{G}_{pr}\tilde{G}_{qs} \ .
\end{equation}
Using this relationship, the extended Fock operator (Eq.~\ref{eq:eff1}) is given by 
\begin{gather}
    \begin{split}
        F_{pr} &= \bar{h}_{pr} -i \sum_{ut} \bar{h}_{pu,rt}\tilde{G}_{tu} \\
            &+ \frac{1}{4}\sum_{ut,vw} \bar{h}_{put,rvw}(\tilde{G}_{vu}\tilde{G}_{wt}-\tilde{G}_{wu}\tilde{G}_{vt}) + \cdots
    \end{split}
\end{gather}
Now, taking the functional derivative of this expression with respect to the equal-time non-interacting Green's function, we have 
\begin{gather}
    \begin{split}~\label{eq:app1}
        i\frac{\delta F_{pr}}{\delta \tilde{G}_{sq}} &=  \bar{h}_{pq,rs} + \frac{i}{4}\sum_{ut,vw} \bar{h}_{put,rwv}(\delta_{vs}\delta_{uq}\tilde{G}_{wt}+\delta_{tq}\delta_{ws}\tilde{G}_{vu}\\
        &-\delta_{ws}\delta_{uq}\tilde{G}_{vt}-\delta_{vs}\delta_{tq}\tilde{G}_{wu}) + \cdots
    \end{split}
\end{gather}
The second term of Eq.~\ref{eq:app1} reduces to four identical terms and combining all together gives 
\begin{gather}
    \begin{split}
        i\frac{\delta F_{pr}}{\delta \tilde{G}_{sq}} &=  \bar{h}_{pq,rs} - i\sum_{ut} \bar{h}_{pqu,rst}\Tilde{G}_{tu}\\
        &+\frac{i}{4}\sum_{tuwv}\bar{h}_{pqtu,rswv}G^{2\p\h,(0)}_{wv,tu}(t-t^+) + \cdots
    \end{split}
\end{gather}
Continuing this procedure for the higher-order matrix elements, the same general structure is observed. The higher-order terms give rise to the general structure of 
\begin{gather}
    \begin{split}
        \frac{1}{(n!)^2}\times n!\times\frac{n}{(n-1)!} = \frac{1}{(n-1!)^2}
    \end{split}
\end{gather}
equivalent contributions. This arises as there are $\frac{1}{(n!)^2}$ terms to be contracted with. The $n$-particle `non-interacting' Green's function gives rise to $n!$ antisymmetrized combinations of independent-particle propagators. The functional derivative of the $n$-particle `non-interacting' Green's function then gives rise to $n$ different terms containing $n-1$ `non-interacting' single-particle Green's functions. These are then regrouped back into `non-interacting' $(n-1)$-particle Green's functions, therefore giving rise to the factor of $\frac{1}{(n-1)!}$ equivalent terms. By application of this formula, the contraction with the two-body term gives one contribution contracted with the `non-interacting' single-particle Green's function, the contraction with the three-body term gives a total contribution of $\frac{1}{4}$ contracted with the `non-interacting' 2ph Green's function, the contraction with four-body term gives a total contribution of $\frac{1}{36}$ contracted with the `non-interacting' 3ph Green's function and so on. As a demonstration, we have
\begin{gather}
    \begin{split}
        &i\frac{\delta}{\delta\tilde{G}_{sq}}\left(-\frac{i}{(3!)^2}\sum_{tuwv\sigma\epsilon}\bar{h}_{ptu\sigma,rwv\epsilon}G^{3\p\h,(0)}_{wv\epsilon,tu\sigma}(t-t^+)\right) =\\
        &\frac{i}{(3!)^2}\times3!\times\frac{3}{2!}\sum_{tuwv}\bar{h}_{pqtu,rswv}G^{2\p\h,(0)}_{wv,tu}(t-t^+)\\
        &=\frac{i}{4}\sum_{tuwv}\bar{h}_{pqtu,rswv}G^{2\p\h,(0)}_{wv,tu}(t-t^+) , 
    \end{split}
\end{gather}
where 
\begin{gather}
    \begin{split}
        -iG^{3\p\h,(0)}_{wv\epsilon,tu\sigma}(t-t^+) = \braket{\Phi_0|a^\dag_ta^\dag_ua^\dag_\sigma a_{\epsilon}a_va_w|\Phi_0} \ .
    \end{split}
\end{gather}
This general relationship holds for the contractions of higher-order matrix elements with `non-interacting' Green's functions. As a result, the final expression is exactly that of the effective two-body interaction, $\chi_{pq,rs}$, and so 
\begin{gather}
    \begin{split}
        i\frac{\delta F_{pr}}{\delta \tilde{G}_{sq}} &= \chi_{pq,rs} \ .
    \end{split}
\end{gather}
The resulting analysis for the two-body effective interaction is identical
\begin{gather}
        \begin{split}
            \chi_{pq,rs} &= \bar{h}_{pq,rs} -i\sum_{tu} \bar{h}_{pqt,rsu}\tilde{G}_{ut} \\
            &+ \frac{i}{4}\sum_{tuwv}\bar{h}_{pqtu,rswv}G^{2\p\h,(0)}_{wv,tu}(t-t^+) + \cdots
        \end{split}
\end{gather}
such that we have 
\begin{gather}
        \begin{split}
            i\frac{\delta \chi_{pq,rs}}{\delta\tilde{G}_{ut}} &= \bar{h}_{pqt,rsu} -i\sum_{tu} \bar{h}_{pqtv,rsuw}\tilde{G}_{wv} \\
            &+ \frac{i}{4}\sum_{vwo\sigma}\bar{h}_{pqtvw,rsuo\sigma}G^{2\p\h,(0)}_{o\sigma,vw}(t-t^+) + \cdots
        \end{split}
\end{gather}
which is exactly the expression for the three-body effective interaction. Therefore
\begin{gather}
        \begin{split}
            i\frac{\delta \chi_{pq,rs}}{\delta\tilde{G}_{ut}} &= \chi_{pqt,rsu} \ .
        \end{split}
\end{gather}
This general relation applies to all effective interactions contained within the similarity transformed Hamiltonian.

\section{Functional derivatives of Goldstone diagrams}~\label{app:gold}
In quantum field theory, contractions are represented by the single-particle Green's function and in coupled-cluster theory these contractions are simply equal-time Green's functions. A contraction is implicitly defined with respect to a many-body state (the reference). As a result, a time-ordered product of two operators may be written in terms of Wick's theorem as 
\begin{equation}
    \mathcal{T}\left\{a_{p}(t_1)a^\dag_{q}(t_2)\right\} = \left\{a_{p}(t_1)a^\dag_{q}(t_2)\right\}_0 + \wick{\c1 a_{p}(t_1) \c1 a^\dag_{q}(t_2)} \ .
\end{equation}
Taking the reference state to be $\ket{\Phi_0}$, the basic contraction is given by the `non-interacting' Green's function as 
\begin{equation}
    \braket{\Phi_0|\mathcal{T}\left\{a_{p}(t_1)a^\dag_{q}(t_2)\right\}|\Phi_0} = \wick{\c1 a_{p}(t_1) \c1 a^\dag_{q}(t_2)} = iG^0_{pq}(t_1,t_2) .
\end{equation}
The only non-zero contractions are where particle/hole creation operators are to the right of particle/hole annihilation operators. Therefore, the only non-zero contractions for particle operators occurs when $t_1>t_2$ such that 
\begin{equation}
    iG^0_{ab}(t_1-t_2) = \wick{\c1 a_{a}(t_1) \c1 a^\dag_{b}(t_2)}
\end{equation}
and for hole operators where $t_2>t_1$ such that 
\begin{equation}
    -iG^0_{ij}(t_1-t_2) = \wick{\c1 a^\dag_{j}(t_2) \c1 a_{i}(t_1)}
\end{equation}
Therefore, the contractions appearing in coupled-cluster theory can be expressed in terms of equal-time particle and hole Green's functions such that 
\begin{gather}
    \begin{split}~\label{eq:contracts}
        iG^0_{ab}(t-t^{-}) &= i\Tilde{G}^{>}_{ab}=\wick{\c1 a_{a} \c1 a^\dag_{b}} = \delta_{ab}\\
        -iG^0_{ij}(t-t^{+}) &= -i\Tilde{G}^{<}_{ij}= \wick{\c1 a^\dag_{j} \c1 a_{i}} = \delta_{ji}
    \end{split}
\end{gather}
where $t^{\pm}$ indicates the evaluation of the time argument from above/below. As a result, the exact ground state correlation energy from coupled-cluster theory may be expressed in terms of equal-time Green's functions (contractions). Now, undoing the contractions in Eq.~\ref{eq:contract} over the occupied and virtual states, using the definition of the `non-interacting' equal-time particle and hole Green's functions in Eq.~\ref{eq:contracts}, we may write the Brueckner correlation energy as 
\begin{equation}~\label{eq:energy}
    E^{\BCC}_{c} = -\frac{1}{4} \sum_{\substack{mnkl\\
    abcd}}\braket{mn||ab}t^{cd}_{kl} \Tilde{G}^{<}_{km}\Tilde{G}^{<}_{ln}\Tilde{G}^{>}_{ac}\Tilde{G}^{>}_{bd} \ .
\end{equation}
Taking the functional derivative of this expression with respect to the equal-time `non-interacting' hole Green's function gives 
\begin{equation}
    i\frac{\delta E^{\BCC}_{c}}{\delta\Tilde{G}^{<}_{ji}} = -\frac{i}{4} \sum_{mnklab}\braket{mn||ab}t^{ab}_{kl} [\delta_{jk}\delta_{mi}\Tilde{G}^{<}_{ln}+\Tilde{G}^{<}_{km}\delta_{jl}\delta_{ni}] \ .
\end{equation}
\begin{figure*}
    \begin{gather*}
\begin{split}
\tilde{\Sigma}^{2\p1\h/2\h1\p}_{pq}(\omega) &= \hspace{2.5mm}\begin{gathered}
\begin{fmfgraph*}(40,40)
    \fmfset{arrow_len}{3mm}
    \fmfleft{i1}
    \fmfright{o1}
    \fmf{zigzag}{o1,i1}
    \fmfv{decor.shape=cross,decor.filled=full, decor.size=1.5thic}{o1}
    \fmfdot{i1}
\end{fmfgraph*}
\end{gathered}\hspace{5mm}+\hspace{5mm}
\begin{gathered}
    \begin{fmfgraph*}(60,60)
    \fmfcurved
    \fmfset{arrow_len}{3mm}
    \fmfleft{i1,i2}
    \fmflabel{}{i1}
    \fmflabel{}{i2}
    \fmfright{o1,o2}
    \fmflabel{}{o1}
    \fmflabel{}{o2}
    \fmf{fermion}{i1,i2}
    \fmf{dbl_zigzag}{o1,i1}
    \fmf{fermion,left=0.3,tension=0}{o1,o2}
    \fmf{dbl_zigzag}{o2,i2}
    \fmf{fermion,left=0.3,tension=0}{o2,o1}
    \fmfdot{o1,o2,i1,i2}
\end{fmfgraph*}
\end{gathered} \hspace{2.5mm}+\hspace{5mm} 
    \begin{gathered}
    \begin{fmfgraph*}(60,80)
    \fmfcurved
    \fmfset{arrow_len}{3mm}
    \fmfleft{i1,i2,i3}
    \fmflabel{}{i1}
    \fmflabel{}{i2}
    \fmfright{o1,o2,o3}
    \fmflabel{}{o1}
    \fmflabel{}{o2}
    \fmf{dbl_zigzag}{i2,o2}
    \fmf{dbl_zigzag}{i3,o3}
    \fmf{dbl_zigzag}{i1,o1}
    \fmf{fermion}{i1,i2}
    \fmf{fermion}{i2,i3}
    \fmf{fermion}{o1,o2}
    \fmf{fermion}{o2,o3}
    \fmf{fermion,left=0.3}{o3,o1}
    \fmfforce{(0.0w,0.0h)}{i1}
    \fmfforce{(1.0w,0.0h)}{o1}
    \fmfforce{(0.0w,1.0h)}{i3}
    \fmfforce{(1.0w,1.0h)}{o3}
    \fmfdot{i1,i2,i3}
    \fmfdot{o1,o2,o3}
\end{fmfgraph*}
\end{gathered}
\hspace{7.5mm}+\hspace{5mm}
\begin{gathered}
\begin{fmfgraph*}(60,80)
    \fmfcurved
    \fmfset{arrow_len}{3mm}
    \fmfleft{i1,i2,i3}
    \fmflabel{}{i1}
    \fmflabel{}{i2}
    \fmfright{o1,o2,o3}
    \fmflabel{}{o1}
    \fmflabel{}{o2}
    \fmf{dbl_zigzag}{i1,v1}
    \fmf{dbl_zigzag}{v1,o1}
    \fmf{dbl_zigzag}{v2,o2}
    \fmf{dbl_zigzag}{i3,v3}
    \fmf{fermion}{i1,i3}
    \fmf{fermion,left=0.3}{o1,o2}
    \fmf{fermion,left=0.3}{o2,o1}
    \fmf{fermion,left=0.3}{v2,v3}
    \fmf{fermion,left=0.3}{v3,v2}
    \fmfforce{(0.0w,0.0h)}{i1}
    \fmfforce{(1.0w,0.0h)}{o1}
    \fmfforce{(0.5w,0.5h)}{v2}
    \fmfforce{(0.5w,0.0h)}{v1}
    \fmfforce{(0.5w,1.0h)}{v3}
    \fmfforce{(0.0w,1.0h)}{i3}
    \fmfforce{(1.0w,1.0h)}{o3}
    \fmfdot{v2,v3}
    \fmfdot{i1,i3}
    \fmfdot{o1,o2}
\end{fmfgraph*}
\end{gathered}\hspace{2.5mm}+\hspace{5mm}
\begin{gathered}
    \begin{fmfgraph*}(60,80)
    \fmfcurved
    \fmfset{arrow_len}{3mm}
    \fmfleft{i1,i2,i3}
    \fmflabel{}{i1}
    \fmflabel{}{i2}
    \fmfright{o1,o2,o3}
    \fmflabel{}{o1}
    \fmflabel{}{o2}
    \fmf{dbl_zigzag}{i1,v1}
    \fmf{dbl_dashes}{i2,v2}
    \fmf{dbl_dashes}{v2,o2}
    \fmf{dbl_zigzag}{i3,v3}
    \fmf{fermion}{i1,i2}
    \fmf{fermion}{v1,v2}
    \fmf{fermion}{o2,v1}
    \fmf{fermion}{v2,v3}
    \fmf{fermion}{v3,o2}
    \fmf{fermion}{i2,i3}
    \fmfforce{(0.0w,0.0h)}{i1}
    \fmfforce{(1.0w,0.0h)}{o1}
    \fmfforce{(0.5w,0.5h)}{v2}
    \fmfforce{(0.5w,0.0h)}{v1}
    \fmfforce{(0.5w,1.0h)}{v3}
    \fmfforce{(0.0w,1.0h)}{i3}
    \fmfforce{(1.0w,1.0h)}{o3}
    \fmfdotn{v}{3}
    \fmfdot{i1,i2,i3}
    \fmfdot{o2}
\end{fmfgraph*}
\end{gathered}\\
\\
&+\hspace{5mm}\begin{gathered}
    \begin{fmfgraph*}(60,60)
    \fmfcurved
    \fmfset{arrow_len}{3mm}
    \fmfleft{i1,i2}
    \fmflabel{}{i1}
    \fmflabel{}{i2}
    \fmfright{o1,o2}
    \fmflabel{}{o1}
    \fmflabel{}{o2}
    \fmf{fermion}{i1,v1}
    \fmf{fermion}{v1,i2}
    \fmf{dbl_zigzag}{o1,i1}
    \fmf{fermion,left=0.3,tension=0}{o1,o2}
    \fmf{dbl_zigzag}{o2,i2}
    \fmf{fermion,left=0.3,tension=0}{o2,o1}
    \fmf{zigzag}{v1,v2}
    \fmfdot{o1,o2,i1,i2,v1}
    \fmfforce{(0.0w,0.0h)}{i1}
    \fmfforce{(0.0w,1.0h)}{i2}
    \fmfforce{(0.0w,0.5h)}{v1}
    \fmfforce{(0.5w,0.5h)}{v2}
     \fmfv{decor.shape=cross,decor.filled=full, decor.size=1.5thic}{v2}
\end{fmfgraph*}
\end{gathered}\hspace{5mm}+\hspace{5mm}
\begin{gathered}
    \begin{fmfgraph*}(60,60)
    \fmfcurved
    \fmfset{arrow_len}{3mm}
    \fmfleft{i1,i2}
    \fmflabel{}{i1}
    \fmflabel{}{i2}
    \fmfright{o1,o2}
    \fmflabel{}{o1}
    \fmflabel{}{o2}
    \fmf{fermion}{i1,i2}
    \fmf{dbl_zigzag}{o1,i1}
    \fmf{fermion,left=0.3,tension=0}{o1,o2}
    \fmf{dbl_zigzag}{o2,i2}
    \fmf{fermion,left=0.3,tension=0}{o2,v1}
     \fmf{fermion,left=0.3,tension=0}{v1,o1}
    \fmf{zigzag}{v1,v2}
    \fmfdot{o1,o2,i1,i2,v1}
    \fmfforce{(1.0w,0.0h)}{o1}
    \fmfforce{(1.0w,1.0h)}{o2}
    \fmfforce{(1.25w,0.5h)}{v1}
    \fmfforce{(1.75w,0.5h)}{v2}
     \fmfv{decor.shape=cross,decor.filled=full, decor.size=1.5thic}{v2}
\end{fmfgraph*}
\end{gathered}\hspace{25mm}+\hspace{5mm}\cdots
\end{split}
\end{gather*}
\caption{Diagrammatic representation of the 1PI CC self-energy diagrams containing 2p1h/2h1p excitation character. The terms depicted are summed to infinite-order by diagonalization of the CC Dyson supermatrix (Eq.~\ref{eq:cc_eom_se_sc}). In this approximation, the interaction matrices are time-ordered, occurring after(before) the initial(final) coupling matrices.}
\label{fig:diag_se}
\end{figure*}
This simplifies to  
\begin{gather}
    \begin{split}
        i\frac{\delta E^{\BCC}_{c}}{\delta\Tilde{G}^{<}_{ji}} &= \frac{1}{4}\sum_{nab} \braket{in||ab}t^{ab}_{jn} + \frac{1}{4}\sum_{mab} \braket{mi||ab}t^{ab}_{mj} \\
        &= \frac{1}{2}\sum_{kab}\braket{ik||ab}t^{ab}_{jk} = \Tilde{\Sigma}^{\infty (0)}_{ij} \ ,
    \end{split}
\end{gather}
where we have used the anti-symmetry of the Coulomb integrals and cluster amplitudes. By the same procedure, taking the functional derivative of Eq.~\ref{eq:energy} with respect to the equal-time `non-interacting' particle Green's function gives 
\begin{equation}
     i\frac{\delta E^{\BCC}_{c}}{\delta\Tilde{G}^{>}_{ba}} =-\frac{1}{2}\sum_{ijc}\braket{ij||ac}t^{bc}_{ij} = \Tilde{\Sigma}^{\infty (0)}_{ab} \ .
\end{equation}
These series of functional derivatives are outlined in Fig.~\ref{fig:SE}. The first-order expression for the BSE kernel may also be obtained by re-writing the coupled-cluster self-energy in Eq.~\ref{eq:se} in terms of the equal-time `non-interacting' particle Green's function as
\begin{gather}
    \begin{split}
        i\frac{\delta\Sigma^{\infty(0)}_{ij}}{\delta\tilde{G}^{>}_{ba}} &= i\frac{\delta}{\delta\tilde{G}^{>}_{ba}} \left(-\frac{1}{2}\sum_{kcdef} \braket{ik||cd}t^{ef}_{jk}\Tilde{G}^{>}_{ce}\Tilde{G}^{>}_{df}\right)\\
        &=\sum_{kc}\braket{ik||bc}t^{ca}_{jk} = \tilde{\Xi}^{c(0)}_{ia,bj} \ .
    \end{split}
\end{gather}
This corresponds to the cutting the particle lines of the coupled-cluster self-energy Goldstone diagram. Finally, we may also derive the 2h1p interaction kernel by taking the functional derivative of the CC-BSE kernel with respect to the equal-time `non-interacting' particle Green's function yielding
\begin{gather}
    \begin{split}
        i\frac{\delta\tilde{\Xi}^{(0)}_{ij,lk}}{\delta\tilde{G}^{>}_{ba}} &= i\frac{\delta}{\delta\tilde{G}^{>}_{ba}}\braket{ij||lk}\\
        &+i\frac{\delta}{\delta\tilde{G}^{>}_{ba}}\left(-\frac{1}{2}\sum_{cdef}\braket{ij||cd}t^{ef}_{kl}\tilde{G}^{>}_{ce}\tilde{G}^{>}_{df}\right)\\
        &=\sum_{c} \braket{ij||bc}t^{ca}_{lk} = \tilde{\chi}^{(0)}_{ija,kbl} \ .
    \end{split}
\end{gather}
This corresponds to taking the third derivative of the Brueckner correlation energy. 

\section{One-particle irreducible coupled-cluster self-energy 2p1h/2h1p excitation diagrams}~\label{app:dyn_ccse}

Here, we present the diagrammatic content of the coupled-cluster self-energy contained in the 2p1h/2h1p excitation character restricted coupled-cluster Dyson supermatrix derived in Section~\ref{sec:rel_eomcc}. The diagrams depicted in Fig.~\ref{fig:diag_se} are the fundamental one-particle irreducible coupled-cluster self-energy diagrams of 2p1h/2h1p excitation character which are subsequently summed to infinite-order by diagonalization of the coupled-cluster Dyson supermatrix. The complete 2h1p/2p1h ISC excitation character coupled-cluster Dyson supermatrix, derived in Section~\ref{sec:rel_eomcc}, is written as
\begin{gather}
    \begin{split}~\label{eq:cc_eom_se_sc}
        &\mathbf{\tilde{D}}^{\text{CC}} = \\
        &\left(\begin{array}{ccc}
            f_{pq}+\tilde{\Sigma}^{\infty}_{pq} & \tilde{U}_{p,iab} & \bar{V}_{p,ija}  \\
            \bar{U}_{jcd,q} & \mathbf{\bar{K}}^{>}_{iab,jcd}+\mathbf{\bar{C}}^{>}_{iab,jcd} & \mathbf{0} \\
             \tilde{V}_{klb,q} & \mathbf{0} &  \mathbf{\bar{K}}^{<}_{ija,klb}+\mathbf{\bar{C}}^{<}_{ija,klb}
        \end{array}\right) ,
    \end{split}
\end{gather}
where the forward- and backward-time coupling matrices are defined as 
\begin{subequations}
    \begin{gather}
        \begin{split}
            \bar{V}_{p,kla} =  \tilde{\Xi}_{pa,kl}
        \end{split}
    \end{gather}
    \begin{gather}
        \begin{split}
            \tilde{V}_{mnb,q} = \tilde{\Xi}_{mn,qb} 
        \end{split}
    \end{gather}
    \begin{gather}
    \begin{split}
        \tilde{U}_{p,kab}  = \tilde{\Xi}_{pk,ab}
    \end{split}
\end{gather}
\begin{gather}
    \begin{split}
        \bar{U}_{lcd,q} = \tilde{\Xi}_{cd,ql} \ .
    \end{split}
\end{gather}
\end{subequations}
Contributions to the forward- and backward-time coupling matrices containing terms such as 
\begin{gather}
    \begin{split}
       \sum_{lb}\tilde{\Xi}_{pl,ka}\frac{\tilde{\Xi}_{ba,lk}}{f_{kk}+f_{ll}-f_{aa}-f_{bb}} 
    \end{split}
\end{gather}
and so on, vanish as a result of the CC amplitude equations: $\tilde{\Xi}_{ba,lk}=0$. Similarly any coupling matrices that contain terms such as $\tilde{\chi}_{abc,ijk}$ also vanish. The backward-time interaction matrices are defined in Eq.~\ref{eq:backward} of the main text and the forward-time interaction matrices between the 2p1h ISCs, $(\mathbf{\bar{K}}^{>}_{iab,jcd}+\mathbf{\bar{C}}^{>}_{iab,jcd})$, are defined as 
\begin{gather}
    \begin{split}~\label{eq:forward}
        (\mathbf{\bar{K}}^{>}_{iab,jcd}+\mathbf{\bar{C}}^{>}_{iab,jcd}) &= \tilde{F}_{ac}\delta_{ij}\delta_{bd}-\tilde{F}_{ad}\delta_{ij}\delta_{bc} + \tilde{F}_{bd}\delta_{ac}\delta_{ij}\\
        &-\tilde{F}_{bc}\delta_{ad}\delta_{ij}- \tilde{F}_{ji}\delta_{ac}\delta_{bd}+\tilde{F}_{ji}\delta_{ad}\delta_{bc}
        \\
        &+\tilde{\Xi}_{ja,ci}\delta_{bd}+\tilde{\Xi}_{jb,di}\delta_{ac}+ \tilde{\Xi}_{ab,cd}\delta_{ij}\\
        &- \tilde{\Xi}_{ja,di}\delta_{bc}- \tilde{\Xi}_{jb,ci}\delta_{ad} + \tilde{\chi}_{jab,cid} \ .
    \end{split}
\end{gather}
Downfolding the coupled-cluster Dyson supermatrix gives the expression for the coupled-cluster self-energy restricted to the coupling to 2p1h/2h1p ISCs over the combined occupied and virtual spin-orbitals
\begin{gather}
    \begin{split}~\label{eq:se_cc}
        &\tilde{\Sigma}^{2\p1\h/2\h1\p}_{pq}(\omega) = \tilde{\Sigma}^{\infty}_{pq} \\
        &+ \frac{1}{4} \sum_{\substack{iab\\jcd}}\tilde{\Xi}_{pi,ab}\left[(\omega+i\eta)\mathbbm{1}-(\mathbf{\bar{K}}^{>}+\mathbf{\bar{C}}^{>})\right]^{-1}_{iab,jcd}\tilde{\Xi}_{cd,qj}\\
        &+ \frac{1}{4} \sum_{\substack{kla\\mnb}}\tilde{\Xi}_{pa,kl}\left[(\omega-i\eta)\mathbbm{1}-(\mathbf{\bar{K}}^{<}+\mathbf{\bar{C}}^{<})\right]^{-1}_{kla,mnb}\tilde{\Xi}_{mn,qb} \ .
    \end{split}
\end{gather}
Diagonalization of the coupled-cluster Dyson supermatrix, Eq.~\ref{eq:cc_eom_se_sc}, gives rise to the infinite-order series of diagrams depicted in Fig.~\ref{fig:diag_se}. The diagrammatic content presented in Fig.~\ref{fig:diag_se} is obtained from Eq.~\ref{eq:se_cc} by using the identity 
\begin{gather}
\begin{split}~\label{eq:schwinger}
    &\left[\omega\mathbbm{1}-(\mathbf{\bar{K}}^{\lessgtr}+\mathbf{\bar{C}}^{\lessgtr})\right]^{-1}_{LL'}=\\
    &\sum_{L''}\left[\omega\mathbbm{1}-\mathbf{\bar{K}}^{\lessgtr}\right]^{-1}_{LL''}\sum_{n=0}^{\infty}\Bigg(\mathbf{\bar{C}}^{\lessgtr}\left[\omega\mathbbm{1}-\mathbf{\bar{K}}^{\lessgtr}\right]^{-1}\Bigg)^{n}_{L''L'}.
\end{split}
\end{gather} 
where $L$ is a general composite index representing either forward- or backward-time ISCs.   
In deriving Eq.~\ref{eq:schwinger}, we define the interaction matrices as
\begin{subequations}
    \begin{align}
    \begin{split}
        \mathbf{\bar{K}}^{>}_{iab,jcd} &= f_{ac}\delta_{ij}\delta_{bd}-f_{ad}\delta_{ij}\delta_{bc} + f_{bd}\delta_{ac}\delta_{ij}\\
        &-f_{bc}\delta_{ad}\delta_{ij}- f_{ji}\delta_{ac}\delta_{bd}+f_{ji}\delta_{ad}\delta_{bc}
    \end{split}\\
    \begin{split}
        \mathbf{\bar{K}}^{<}_{ija,klb} &=   f_{ik}\delta_{ab}\delta_{jl}+f_{jl}\delta_{ab}\delta_{ik}-f_{ba}\delta_{ik}\delta_{jl}\\
    &-f_{il}\delta_{ab}\delta_{jk}-f_{jk}\delta_{ab}\delta_{il}+f_{ba}\delta_{il}\delta_{jk}
    \end{split}
\end{align}
\end{subequations}
with the remaining terms in Eq.~\ref{eq:forward}/Eq.~\ref{eq:backward} constituting the definition of the $\mathbf{\bar{C}}^{>}_{iab,jcd}/\mathbf{\bar{C}}^{<}_{ija,klb}$ matrices.

The diagrammatic notation depicted in Fig.~\ref{fig:diag_se} is as follows. The first diagram is the static CC self-energy, $\tilde{\Sigma}^{\infty}_{pq}$ and is depicted diagrammatically in Eq.~\ref{eq:dia_stat}. 
\begin{figure}[ht]
\begin{gather}
    \begin{split}~\label{eq:dia_stat}
        \tilde{\Sigma}^{\infty}_{pq} =\hspace{2.5mm}
        \begin{gathered}\begin{fmfgraph*}(30,30)
    \fmfset{arrow_len}{3mm}
    \fmfleft{i1}
    \fmfright{o1}
    \fmf{zigzag}{o1,i1}
    \fmfv{decor.shape=cross,decor.filled=full, decor.size=1.5thic}{o1}
    \fmfdot{i1}
\end{fmfgraph*}
\end{gathered} \hspace{5mm}&=\hspace{5mm} \begin{gathered}
\begin{fmfgraph*}(30,30)
    \fmfset{arrow_len}{3mm}
    \fmfleft{i1,i2,i3}
    \fmfright{o1,o2,o3}
    \fmf{fermion}{i1,i2}
    \fmf{fermion}{i2,i3}
    \fmf{dbl_dashes}{i2,o2}
    \fmfforce{(0.0w,0.h)}{i1}
    \fmfforce{(0.0w,0.5h)}{i2}
    \fmfforce{(0.0w,1.0h)}{i3}
    \fmfdot{i2}
    \fmfv{decor.shape=cross,decor.filled=full, decor.size=1.5thic}{o2}
\end{fmfgraph*}
\end{gathered}\hspace{5mm}+\hspace{7.5mm}
\begin{gathered}
    \begin{fmfgraph*}(30,30)
    \fmfcurved
    \fmfset{arrow_len}{3mm}
    \fmfleft{i1,i2}
    \fmflabel{}{i1}
    \fmflabel{}{i2}
    \fmfright{o1,o2}
    \fmflabel{}{o1}
    \fmflabel{}{o2}
    \fmf{dbl_wiggly}{i1,o1}
    \fmf{fermion,left=0.3,tension=0}{o1,o2}
    \fmf{fermion,left=0.3,tension=0}{o2,o1}
    \fmf{dbl_plain}{v1,o2}
    \fmf{dbl_plain}{o2,v2}
    \fmf{fermion}{v3,i1}
    \fmf{fermion}{i1,v4}
    \fmfforce{(1.05w,1.0h)}{v2}
    \fmfforce{(0.75w,1.0h)}{v1}
    \fmfforce{(0.0w,-0.5h)}{v3}
    \fmfforce{(0.0w,0.5h)}{v4}
    \fmfforce{(0.0w,0.0h)}{i1}
    \fmfdot{o1,i1}
\end{fmfgraph*}
\end{gathered}\\
&+\hspace{5mm}
\begin{gathered}
    \begin{fmfgraph*}(40,30)
    \fmfcurved
    \fmfset{arrow_len}{3mm}
    \fmfleft{i1,i2}
    \fmflabel{}{i1}
    \fmflabel{}{i2}
    \fmfright{o1,o2}
    \fmflabel{}{o1}
    \fmflabel{}{o2}
    \fmf{dashes}{o1,v1}
    \fmf{dashes}{i1,v1}
    \fmf{fermion,left=0.3,tension=0}{o1,o2}
    \fmf{fermion,left=0.3,tension=0}{v1,v2}
    \fmf{fermion,left=0.3,tension=0}{v2,v1}
    \fmf{phantom}{v2,i2}
    \fmf{dbl_plain}{o2,v2}
    \fmf{fermion,left=0.3,tension=0}{o2,o1}
    \fmf{fermion}{v3,i1}
    \fmf{fermion}{i1,v4}
    \fmfdot{o1,i1,v1}
    \fmfforce{(0.0w,-0.5h)}{v3}
    \fmfforce{(0.0w,0.5h)}{v4}
    \fmfforce{(0.0w,0.0h)}{i1}
    \fmfforce{(0.5w,1.0h)}{v2}
    \fmfforce{(0.5w,0.0h)}{v1}
\end{fmfgraph*}
\end{gathered}\hspace{5mm}+\hspace{2.5mm}\cdots 
    \end{split}
\end{gather}
\end{figure}

The two-body interaction is the CC-BSE kernel, $\tilde{\Xi}_{pq,rs}$ and is represented by the following diagrammatic series in Eq.~\ref{eq:bse_diag}:
\begin{figure}[ht]
\begin{gather}
        \begin{split}~\label{eq:bse_diag}
            \tilde{\Xi}_{pq,rs} =\hspace{2.5mm}
            \begin{gathered}
    \begin{fmfgraph*}(30,30)
    \fmfset{arrow_len}{3mm}
    \fmfleft{i1}
    \fmfright{o1}
    \fmf{dbl_zigzag}{o1,i1}
    \fmfdot{i1,o1}
\end{fmfgraph*}
\end{gathered} \hspace{2.5mm} &= \hspace{2.5mm}
\begin{gathered}
    \begin{fmfgraph*}(30,30)
    \fmfset{arrow_len}{3mm}
    \fmfleft{i1}
    \fmfright{o1}
    \fmf{dbl_wiggly}{o1,i1}
    \fmfdot{i1,o1}
\end{fmfgraph*}
\end{gathered}\hspace{5mm}+\hspace{7.5mm}
\begin{gathered}
    \begin{fmfgraph*}(40,30)
    \fmfcurved
    \fmfset{arrow_len}{3mm}
    \fmfleft{i1,i2}
    \fmflabel{}{i1}
    \fmflabel{}{i2}
    \fmfright{o1,o2}
    \fmflabel{}{o1}
    \fmflabel{}{o2}
    \fmf{dashes}{o1,v1}
    \fmf{dashes}{i1,v1}
    \fmf{fermion,left=0.3,tension=0}{o1,o2}
    \fmf{phantom}{v2,i2}
    \fmf{fermion,left=0.3,tension=0}{o1,o2}
    \fmf{dbl_plain}{v3,o2}
    \fmf{dbl_plain}{o2,v4}
    \fmf{fermion,left=0.3,tension=0}{o2,o1}
    \fmf{fermion,left=0.3,tension=0}{o2,o1}
    \fmfdot{o1,i1,v1}
    \fmfforce{(1.05w,1.0h)}{v3}
    \fmfforce{(0.75w,1.0h)}{v4}
\end{fmfgraph*}
\end{gathered}\\
&+\hspace{5mm}
\begin{gathered}
    \begin{fmfgraph*}(40,30)
    \fmfcurved
    \fmfset{arrow_len}{3mm}
    \fmfleft{i1,i2}
    \fmflabel{}{i1}
    \fmflabel{}{i2}
    \fmfright{o1,o2}
    \fmflabel{}{o1}
    \fmflabel{}{o2}
    \fmf{dbl_dashes}{v1,i1}
    \fmf{dbl_dashes}{o1,v1}
    \fmf{dbl_dashes}{v1,v3}
    \fmf{dbl_dashes}{v3,o1}
    \fmf{dbl_plain}{v4,o2}
    \fmf{fermion,left=0.3,tension=0}{o1,o2}
    \fmf{fermion,left=0.3,tension=0}{v3,v4}
    \fmf{fermion,left=0.3,tension=0}{v4,v3}
    \fmf{fermion,left=0.3,tension=0}{o2,o1}
    \fmfdot{o1,i1,v1,v3}
    \fmfforce{(0.0w,0.0h)}{i1}
    \fmfforce{(1.0w,0.0h)}{o1}
    \fmfforce{(0.25w,1h)}{v2}
    \fmfforce{(0.25w,0.0h)}{v1}
    \fmfforce{(0.625w,0.0h)}{v3}
    \fmfforce{(0.625w,1.0h)}{v4}
    \fmfforce{(0.0w,1.0h)}{i2}
    \fmfforce{(1.0w,1.0h)}{o2}
\end{fmfgraph*}
\end{gathered}\hspace{5mm}+\hspace{2.5mm} \cdots  
        \end{split}
    \end{gather}
\end{figure}

The three-body CC effective interaction, $\tilde{\chi}_{pqr,stu}$, corresponding to the kernel of the 2h1p/2p1h Green's function is depicted in Eq.~\ref{eq:3-bod}.
\begin{figure}[ht]
\begin{gather}
        \begin{split}~\label{eq:3-bod}
            \tilde{\chi}_{pqr,stu}=\hspace{2.5mm}\begin{gathered}
    \begin{fmfgraph*}(30,30)
    \fmfset{arrow_len}{3mm}
    \fmfleft{i1}
    \fmfright{o1}
    \fmf{dbl_dashes}{i1,v1}
    \fmf{dbl_dashes}{v1,o1}
    \fmfdot{i1,o1,v1}
\end{fmfgraph*}
\end{gathered}\hspace{5mm}&=\hspace{5mm}\begin{gathered}
    \begin{fmfgraph*}(30,30)
    \fmfset{arrow_len}{3mm}
    \fmfleft{i1}
    \fmfright{o1}
    \fmf{dashes}{i1,v1}
    \fmf{dashes}{v1,o1}
    \fmfdot{i1,o1,v1}
\end{fmfgraph*}
\end{gathered}\hspace{5mm}+\hspace{5mm}
\begin{gathered}
    \begin{fmfgraph*}(40,30)
    \fmfcurved
    \fmfset{arrow_len}{3mm}
    \fmfleft{i1,i2}
    \fmflabel{}{i1}
    \fmflabel{}{i2}
    \fmfright{o1,o2}
    \fmflabel{}{o1}
    \fmflabel{}{o2}
    \fmf{dbl_dashes}{i1,v1}
    \fmf{dbl_dashes}{v1,v5}
    \fmf{dbl_dashes}{v5,o1}
    \fmf{fermion,left=0.3,tension=0}{o1,o2}
    \fmf{phantom}{v2,i2}
    \fmf{fermion,left=0.3,tension=0}{o1,o2}
    \fmf{dbl_plain}{v3,o2}
    \fmf{dbl_plain}{o2,v4}
    \fmf{fermion,left=0.3,tension=0}{o2,o1}
    \fmf{fermion,left=0.3,tension=0}{o2,o1}
    \fmfdot{o1,i1,v1,v5}
    \fmfforce{(1.05w,1.0h)}{v3}
    \fmfforce{(0.75w,1.0h)}{v4}
    \fmfforce{(0.67w,0.0h)}{v5}
    \fmfforce{(0.33w,0.0h)}{v1}
\end{fmfgraph*}
\end{gathered}\\
&+\hspace{5mm}
\begin{gathered}
    \begin{fmfgraph*}(40,30)
    \fmfcurved
    \fmfset{arrow_len}{3mm}
    \fmfleft{i1,i2}
    \fmflabel{}{i1}
    \fmflabel{}{i2}
    \fmfright{o1,o2}
    \fmflabel{}{o1}
    \fmflabel{}{o2}
    \fmf{dbl_dashes}{i1,v1}
    \fmf{dbl_dashes}{v1,v5}
    \fmf{dbl_dashes}{v5,v3}
    \fmf{dbl_dashes}{v3,o1}
    \fmf{dbl_plain}{v4,o2}
    \fmf{fermion,left=0.3,tension=0}{o1,o2}
    \fmf{fermion,left=0.3,tension=0}{v3,v4}
    \fmf{fermion,left=0.3,tension=0}{v4,v3}
    \fmf{fermion,left=0.3,tension=0}{o2,o1}
    \fmfdot{o1,i1,v1,v3,v5}
    \fmfforce{(0.0w,0.0h)}{i1}
    \fmfforce{(1.0w,0.0h)}{o1}
    \fmfforce{(0.25w,1h)}{v2}
    \fmfforce{(0.25w,0.0h)}{v1}
    \fmfforce{(0.625w,0.0h)}{v3}
    \fmfforce{(0.625w,1.0h)}{v4}
    \fmfforce{(0.0w,1.0h)}{i2}
    \fmfforce{(1.0w,1.0h)}{o2}
    \fmfforce{(0.425w,0.0h)}{v5}
\end{fmfgraph*}
\end{gathered}\hspace{5mm}+\hspace{2.5mm} \cdots
        \end{split}
    \end{gather}
    \end{figure} 

The diagrams presented in Eqs~\ref{eq:dia_stat},~\ref{eq:bse_diag} and~\ref{eq:3-bod} are exactly the interaction kernels derived in Sections~\ref{sec:sim_mbgf} and~\ref{sec:BSE} of the main text. 

Fig.~\ref{fig:diag_se} translates Eq.~\ref{eq:se_cc} into the fundamental 1PI CC self-energy diagrams that are encoded by diagonalization of Eq.~\ref{eq:cc_eom_se_sc}. The series in Fig.~\ref{fig:diag_se} continues through to infinite-order containing all combinations of insertions of the different dynamical contributions as the non-interacting Green's function lines are subsequently dressed by self-energy insertions. Through the analysis presented here, we can view the diagrammatic content contained in the CC Dyson supermatrix in terms of the Algebraic Diagrammatic Construction method.~\cite{schirmer1983new,schirmer2018many,raimondi2018algebraic}

\end{fmffile}

\bibliography{cc_se}

\begin{thebibliography}{58}%
\makeatletter
\providecommand \@ifxundefined [1]{%
 \@ifx{#1\undefined}
}%
\providecommand \@ifnum [1]{%
 \ifnum #1\expandafter \@firstoftwo
 \else \expandafter \@secondoftwo
 \fi
}%
\providecommand \@ifx [1]{%
 \ifx #1\expandafter \@firstoftwo
 \else \expandafter \@secondoftwo
 \fi
}%
\providecommand \natexlab [1]{#1}%
\providecommand \enquote  [1]{``#1''}%
\providecommand \bibnamefont  [1]{#1}%
\providecommand \bibfnamefont [1]{#1}%
\providecommand \citenamefont [1]{#1}%
\providecommand \href@noop [0]{\@secondoftwo}%
\providecommand \href [0]{\begingroup \@sanitize@url \@href}%
\providecommand \@href[1]{\@@startlink{#1}\@@href}%
\providecommand \@@href[1]{\endgroup#1\@@endlink}%
\providecommand \@sanitize@url [0]{\catcode `\\12\catcode `\$12\catcode `\&12\catcode `\#12\catcode `\^12\catcode `\_12\catcode `\%12\relax}%
\providecommand \@@startlink[1]{}%
\providecommand \@@endlink[0]{}%
\providecommand \url  [0]{\begingroup\@sanitize@url \@url }%
\providecommand \@url [1]{\endgroup\@href {#1}{\urlprefix }}%
\providecommand \urlprefix  [0]{URL }%
\providecommand \Eprint [0]{\href }%
\providecommand \doibase [0]{http://dx.doi.org/}%
\providecommand \selectlanguage [0]{\@gobble}%
\providecommand \bibinfo  [0]{\@secondoftwo}%
\providecommand \bibfield  [0]{\@secondoftwo}%
\providecommand \translation [1]{[#1]}%
\providecommand \BibitemOpen [0]{}%
\providecommand \bibitemStop [0]{}%
\providecommand \bibitemNoStop [0]{.\EOS\space}%
\providecommand \EOS [0]{\spacefactor3000\relax}%
\providecommand \BibitemShut  [1]{\csname bibitem#1\endcsname}%
\let\auto@bib@innerbib\@empty
\bibitem [{\citenamefont {Fetter}\ and\ \citenamefont {Walecka}(1971)}]{Quantum}%
  \BibitemOpen
  \bibfield  {author} {\bibinfo {author} {\bibfnamefont {A}~\bibnamefont {Fetter}}\ and\ \bibinfo {author} {\bibfnamefont {J~D}\ \bibnamefont {Walecka}},\ }\href@noop {} {\emph {\bibinfo {title} {Quantum theory of many particle systems}}}\ (\bibinfo  {publisher} {McGraw-Hill},\ \bibinfo {year} {1971})\BibitemShut {NoStop}%
\bibitem [{\citenamefont {Mahan}(2000)}]{mahan2000many}%
  \BibitemOpen
  \bibfield  {author} {\bibinfo {author} {\bibfnamefont {G~D}\ \bibnamefont {Mahan}},\ }\href@noop {} {\emph {\bibinfo {title} {Many-particle physics}}}\ (\bibinfo  {publisher} {Springer Science \& Business Media},\ \bibinfo {year} {2000})\BibitemShut {NoStop}%
\bibitem [{\citenamefont {Stefanucci}\ and\ \citenamefont {Van~Leeuwen}(2013)}]{stefanucci2013nonequilibrium}%
  \BibitemOpen
  \bibfield  {author} {\bibinfo {author} {\bibfnamefont {G}~\bibnamefont {Stefanucci}}\ and\ \bibinfo {author} {\bibfnamefont {R}~\bibnamefont {Van~Leeuwen}},\ }\href@noop {} {\emph {\bibinfo {title} {Nonequilibrium many-body theory of quantum systems: a modern introduction}}}\ (\bibinfo  {publisher} {Cambridge University Press},\ \bibinfo {year} {2013})\BibitemShut {NoStop}%
\bibitem [{\citenamefont {Dyson}(1949)}]{dyson1949radiation}%
  \BibitemOpen
  \bibfield  {author} {\bibinfo {author} {\bibfnamefont {F~J}\ \bibnamefont {Dyson}},\ }\bibfield  {title} {\enquote {\bibinfo {title} {The radiation theories of {T}omonaga, {S}chwinger and {F}eynman},}\ }\href {https://doi.org/10.1103/PhysRev.75.486} {\bibfield  {journal} {\bibinfo  {journal} {Physical Review}\ }\textbf {\bibinfo {volume} {75}},\ \bibinfo {pages} {486} (\bibinfo {year} {1949})}\BibitemShut {NoStop}%
\bibitem [{\citenamefont {Hedin}(1965)}]{hedin1965new}%
  \BibitemOpen
  \bibfield  {author} {\bibinfo {author} {\bibfnamefont {L}~\bibnamefont {Hedin}},\ }\bibfield  {title} {\enquote {\bibinfo {title} {New method for calculating the one-particle {G}reen's function with application to the electron-gas problem},}\ }\href {https://doi.org/10.1103/PhysRev.139.A796} {\bibfield  {journal} {\bibinfo  {journal} {Physical Review}\ }\textbf {\bibinfo {volume} {139}},\ \bibinfo {pages} {A796} (\bibinfo {year} {1965})}\BibitemShut {NoStop}%
\bibitem [{\citenamefont {Pokhilko}\ \emph {et~al.}(2021)\citenamefont {Pokhilko}, \citenamefont {Iskakov}, \citenamefont {Yeh},\ and\ \citenamefont {Zgid}}]{pokhilko2021evaluation}%
  \BibitemOpen
  \bibfield  {author} {\bibinfo {author} {\bibfnamefont {P}~\bibnamefont {Pokhilko}}, \bibinfo {author} {\bibfnamefont {S}~\bibnamefont {Iskakov}}, \bibinfo {author} {\bibfnamefont {C-N}\ \bibnamefont {Yeh}}, \ and\ \bibinfo {author} {\bibfnamefont {D}~\bibnamefont {Zgid}},\ }\bibfield  {title} {\enquote {\bibinfo {title} {Evaluation of two-particle properties within finite-temperature self-consistent one-particle {G}reen\textquotesingle s function methods: Theory and application to {$GW$} and {GF2}},}\ }\href {https://doi.org/10.1063/5.0054661} {\bibfield  {journal} {\bibinfo  {journal} {The Journal of Chemical Physics}\ }\textbf {\bibinfo {volume} {155}},\ \bibinfo {pages} {024119} (\bibinfo {year} {2021})}\BibitemShut {NoStop}%
\bibitem [{\citenamefont {Martin}\ \emph {et~al.}(2016)\citenamefont {Martin}, \citenamefont {Reining},\ and\ \citenamefont {Ceperley}}]{martin2016interacting}%
  \BibitemOpen
  \bibfield  {author} {\bibinfo {author} {\bibfnamefont {R~M}\ \bibnamefont {Martin}}, \bibinfo {author} {\bibfnamefont {L}~\bibnamefont {Reining}}, \ and\ \bibinfo {author} {\bibfnamefont {D~M}\ \bibnamefont {Ceperley}},\ }\href@noop {} {\emph {\bibinfo {title} {Interacting electrons}}}\ (\bibinfo  {publisher} {Cambridge University Press},\ \bibinfo {year} {2016})\BibitemShut {NoStop}%
\bibitem [{\citenamefont {Verdozzi}\ \emph {et~al.}(1995)\citenamefont {Verdozzi}, \citenamefont {Godby},\ and\ \citenamefont {Holloway}}]{verdozzi1995evaluation}%
  \BibitemOpen
  \bibfield  {author} {\bibinfo {author} {\bibfnamefont {C}~\bibnamefont {Verdozzi}}, \bibinfo {author} {\bibfnamefont {R~W}\ \bibnamefont {Godby}}, \ and\ \bibinfo {author} {\bibfnamefont {S}~\bibnamefont {Holloway}},\ }\bibfield  {title} {\enquote {\bibinfo {title} {Evaluation of {$GW$} approximations for the self-energy of a {H}ubbard cluster},}\ }\href {https://doi.org/10.1103/PhysRevLett.74.2327} {\bibfield  {journal} {\bibinfo  {journal} {Physical Review Letters}\ }\textbf {\bibinfo {volume} {74}},\ \bibinfo {pages} {2327} (\bibinfo {year} {1995})}\BibitemShut {NoStop}%
\bibitem [{\citenamefont {Von~Friesen}\ \emph {et~al.}(2009)\citenamefont {Von~Friesen}, \citenamefont {Verdozzi},\ and\ \citenamefont {Almbladh}}]{von2009successes}%
  \BibitemOpen
  \bibfield  {author} {\bibinfo {author} {\bibfnamefont {M~P}\ \bibnamefont {Von~Friesen}}, \bibinfo {author} {\bibfnamefont {C}~\bibnamefont {Verdozzi}}, \ and\ \bibinfo {author} {\bibfnamefont {C-O}\ \bibnamefont {Almbladh}},\ }\bibfield  {title} {\enquote {\bibinfo {title} {Successes and failures of {Kadanoff-Baym} dynamics in {Hubbard} nanoclusters},}\ }\href {https://doi.org/10.1103/PhysRevLett.103.176404} {\bibfield  {journal} {\bibinfo  {journal} {Physical Review Letters}\ }\textbf {\bibinfo {volume} {103}},\ \bibinfo {pages} {176404} (\bibinfo {year} {2009})}\BibitemShut {NoStop}%
\bibitem [{\citenamefont {Romaniello}\ \emph {et~al.}(2009)\citenamefont {Romaniello}, \citenamefont {Guyot},\ and\ \citenamefont {Reining}}]{romaniello2009self}%
  \BibitemOpen
  \bibfield  {author} {\bibinfo {author} {\bibfnamefont {P}~\bibnamefont {Romaniello}}, \bibinfo {author} {\bibfnamefont {S}~\bibnamefont {Guyot}}, \ and\ \bibinfo {author} {\bibfnamefont {L}~\bibnamefont {Reining}},\ }\bibfield  {title} {\enquote {\bibinfo {title} {The self-energy beyond {$GW$}: Local and nonlocal vertex corrections},}\ }\href {https://doi.org/10.1063/1.3249965} {\bibfield  {journal} {\bibinfo  {journal} {The Journal of Chemical Physics}\ }\textbf {\bibinfo {volume} {131}},\ \bibinfo {pages} {154111} (\bibinfo {year} {2009})}\BibitemShut {NoStop}%
\bibitem [{\citenamefont {von Friesen}\ \emph {et~al.}(2010)\citenamefont {von Friesen}, \citenamefont {Verdozzi},\ and\ \citenamefont {Almbladh}}]{von2010kadanoff}%
  \BibitemOpen
  \bibfield  {author} {\bibinfo {author} {\bibfnamefont {M~P}\ \bibnamefont {von Friesen}}, \bibinfo {author} {\bibfnamefont {C}~\bibnamefont {Verdozzi}}, \ and\ \bibinfo {author} {\bibfnamefont {C-O}\ \bibnamefont {Almbladh}},\ }\bibfield  {title} {\enquote {\bibinfo {title} {{Kadanoff-Baym} dynamics of {Hubbard} clusters: Performance of many-body schemes, correlation-induced damping and multiple steady and quasi-steady states},}\ }\href {http://dx.doi.org/10.1103/PhysRevB.82.155108} {\bibfield  {journal} {\bibinfo  {journal} {Physical Review B}\ }\textbf {\bibinfo {volume} {82}},\ \bibinfo {pages} {155108} (\bibinfo {year} {2010})}\BibitemShut {NoStop}%
\bibitem [{\citenamefont {Romaniello}\ \emph {et~al.}(2012)\citenamefont {Romaniello}, \citenamefont {Bechstedt},\ and\ \citenamefont {Reining}}]{romaniello2012beyond}%
  \BibitemOpen
  \bibfield  {author} {\bibinfo {author} {\bibfnamefont {P}~\bibnamefont {Romaniello}}, \bibinfo {author} {\bibfnamefont {F}~\bibnamefont {Bechstedt}}, \ and\ \bibinfo {author} {\bibfnamefont {L}~\bibnamefont {Reining}},\ }\bibfield  {title} {\enquote {\bibinfo {title} {Beyond the {$GW$} approximation: Combining correlation channels},}\ }\href {https://doi.org/10.1103/PhysRevB.85.155131} {\bibfield  {journal} {\bibinfo  {journal} {Physical Review B}\ }\textbf {\bibinfo {volume} {85}},\ \bibinfo {pages} {155131} (\bibinfo {year} {2012})}\BibitemShut {NoStop}%
\bibitem [{\citenamefont {Kohn}\ and\ \citenamefont {Luttinger}(1960)}]{kohn1960ground}%
  \BibitemOpen
  \bibfield  {author} {\bibinfo {author} {\bibfnamefont {W}~\bibnamefont {Kohn}}\ and\ \bibinfo {author} {\bibfnamefont {J~M}\ \bibnamefont {Luttinger}},\ }\bibfield  {title} {\enquote {\bibinfo {title} {Ground-state energy of a many-fermion system},}\ }\href {https://doi.org/10.1103/PhysRev.118.41} {\bibfield  {journal} {\bibinfo  {journal} {Physical Review}\ }\textbf {\bibinfo {volume} {118}},\ \bibinfo {pages} {41} (\bibinfo {year} {1960})}\BibitemShut {NoStop}%
\bibitem [{\citenamefont {Luttinger}\ and\ \citenamefont {Ward}(1960)}]{luttinger1960ground}%
  \BibitemOpen
  \bibfield  {author} {\bibinfo {author} {\bibfnamefont {J~M}\ \bibnamefont {Luttinger}}\ and\ \bibinfo {author} {\bibfnamefont {J~C}\ \bibnamefont {Ward}},\ }\bibfield  {title} {\enquote {\bibinfo {title} {Ground-state energy of a many-fermion system. {II}},}\ }\href {https://doi.org/10.1103/PhysRev.118.1417} {\bibfield  {journal} {\bibinfo  {journal} {Physical Review}\ }\textbf {\bibinfo {volume} {118}},\ \bibinfo {pages} {1417} (\bibinfo {year} {1960})}\BibitemShut {NoStop}%
\bibitem [{\citenamefont {Luttinger}(1961)}]{luttinger1961analytic}%
  \BibitemOpen
  \bibfield  {author} {\bibinfo {author} {\bibfnamefont {J~M}\ \bibnamefont {Luttinger}},\ }\bibfield  {title} {\enquote {\bibinfo {title} {Analytic properties of single-particle propagators for many-fermion systems},}\ }\href {https://doi.org/10.1103/PhysRev.121.942} {\bibfield  {journal} {\bibinfo  {journal} {Physical Review}\ }\textbf {\bibinfo {volume} {121}},\ \bibinfo {pages} {942} (\bibinfo {year} {1961})}\BibitemShut {NoStop}%
\bibitem [{\citenamefont {Potthoff}(2014)}]{potthoff2014making}%
  \BibitemOpen
  \bibfield  {author} {\bibinfo {author} {\bibfnamefont {M}~\bibnamefont {Potthoff}},\ }\href@noop {} {\emph {\bibinfo {title} {Making use of self-energy functionals: the variational cluster approximation}}}\ (\bibinfo  {publisher} {e-print},\ \bibinfo {year} {2014})\BibitemShut {NoStop}%
\bibitem [{\citenamefont {van Setten}\ \emph {et~al.}(2013)\citenamefont {van Setten}, \citenamefont {Weigend},\ and\ \citenamefont {Evers}}]{van2013gw}%
  \BibitemOpen
  \bibfield  {author} {\bibinfo {author} {\bibfnamefont {M~J}\ \bibnamefont {van Setten}}, \bibinfo {author} {\bibfnamefont {F}~\bibnamefont {Weigend}}, \ and\ \bibinfo {author} {\bibfnamefont {F}~\bibnamefont {Evers}},\ }\bibfield  {title} {\enquote {\bibinfo {title} {The {$GW$}-method for quantum chemistry applications: Theory and implementation},}\ }\href {https://doi.org/10.1021/ct300648t} {\bibfield  {journal} {\bibinfo  {journal} {Journal of Chemical Theory and Computation}\ }\textbf {\bibinfo {volume} {9}},\ \bibinfo {pages} {232--246} (\bibinfo {year} {2013})}\BibitemShut {NoStop}%
\bibitem [{\citenamefont {Nooijen}\ and\ \citenamefont {Snijders}(1992)}]{nooijen1992coupled}%
  \BibitemOpen
  \bibfield  {author} {\bibinfo {author} {\bibfnamefont {M}~\bibnamefont {Nooijen}}\ and\ \bibinfo {author} {\bibfnamefont {J~G}\ \bibnamefont {Snijders}},\ }\bibfield  {title} {\enquote {\bibinfo {title} {{Coupled}-cluster approach to the single-particle {Green's} function},}\ }\href {https://doi.org/10.1002/qua.560440808} {\bibfield  {journal} {\bibinfo  {journal} {International Journal of Quantum Chemistry}\ }\textbf {\bibinfo {volume} {44}},\ \bibinfo {pages} {55--83} (\bibinfo {year} {1992})}\BibitemShut {NoStop}%
\bibitem [{\citenamefont {Nooijen}\ and\ \citenamefont {Snijders}(1993)}]{nooijen1993coupled}%
  \BibitemOpen
  \bibfield  {author} {\bibinfo {author} {\bibfnamefont {M}~\bibnamefont {Nooijen}}\ and\ \bibinfo {author} {\bibfnamefont {J~G}\ \bibnamefont {Snijders}},\ }\bibfield  {title} {\enquote {\bibinfo {title} {{Coupled}-cluster {Green's} function method: Working equations and applications},}\ }\href {https://doi.org/10.1002/qua.560480103} {\bibfield  {journal} {\bibinfo  {journal} {International Journal of Quantum Chemistry}\ }\textbf {\bibinfo {volume} {48}},\ \bibinfo {pages} {15--48} (\bibinfo {year} {1993})}\BibitemShut {NoStop}%
\bibitem [{\citenamefont {Peng}\ and\ \citenamefont {Kowalski}(2018)}]{peng2018green}%
  \BibitemOpen
  \bibfield  {author} {\bibinfo {author} {\bibfnamefont {B}~\bibnamefont {Peng}}\ and\ \bibinfo {author} {\bibfnamefont {K}~\bibnamefont {Kowalski}},\ }\bibfield  {title} {\enquote {\bibinfo {title} {Green’s function coupled-cluster approach: Simulating photoelectron spectra for realistic molecular systems},}\ }\href {https://doi.org/10.1021/acs.jctc.8b00313} {\bibfield  {journal} {\bibinfo  {journal} {Journal of Chemical Theory and Computation}\ }\textbf {\bibinfo {volume} {14}},\ \bibinfo {pages} {4335--4352} (\bibinfo {year} {2018})}\BibitemShut {NoStop}%
\bibitem [{\citenamefont {Shee}\ and\ \citenamefont {Zgid}(2019)}]{shee2019coupled}%
  \BibitemOpen
  \bibfield  {author} {\bibinfo {author} {\bibfnamefont {A}~\bibnamefont {Shee}}\ and\ \bibinfo {author} {\bibfnamefont {D}~\bibnamefont {Zgid}},\ }\bibfield  {title} {\enquote {\bibinfo {title} {Coupled-cluster as an impurity solver for {G}reen’s function embedding methods},}\ }\href {https://doi.org/10.1021/acs.jctc.9b00603} {\bibfield  {journal} {\bibinfo  {journal} {Journal of Chemical Theory and Computation}\ }\textbf {\bibinfo {volume} {15}},\ \bibinfo {pages} {6010--6024} (\bibinfo {year} {2019})}\BibitemShut {NoStop}%
\bibitem [{\citenamefont {Backhouse}\ and\ \citenamefont {Booth}(2022)}]{backhouse2022constructing}%
  \BibitemOpen
  \bibfield  {author} {\bibinfo {author} {\bibfnamefont {O~J}\ \bibnamefont {Backhouse}}\ and\ \bibinfo {author} {\bibfnamefont {G~H}\ \bibnamefont {Booth}},\ }\bibfield  {title} {\enquote {\bibinfo {title} {Constructing “full-frequency” spectra via moment constraints for coupled-cluster {G}reen’s functions},}\ }\href {https://doi.org/10.1021/acs.jctc.2c00670} {\bibfield  {journal} {\bibinfo  {journal} {Journal of Chemical Theory and Computation}\ }\textbf {\bibinfo {volume} {18}},\ \bibinfo {pages} {6622--6636} (\bibinfo {year} {2022})}\BibitemShut {NoStop}%
\bibitem [{\citenamefont {Bintrim}\ and\ \citenamefont {Berkelbach}(2021)}]{bintrim2021full}%
  \BibitemOpen
  \bibfield  {author} {\bibinfo {author} {\bibfnamefont {S~J}\ \bibnamefont {Bintrim}}\ and\ \bibinfo {author} {\bibfnamefont {T~C}\ \bibnamefont {Berkelbach}},\ }\bibfield  {title} {\enquote {\bibinfo {title} {Full-frequency {$GW$} without frequency},}\ }\href {https://doi.org/10.1063/5.0035141} {\bibfield  {journal} {\bibinfo  {journal} {The Journal of Chemical Physics}\ }\textbf {\bibinfo {volume} {154}},\ \bibinfo {pages} {041101} (\bibinfo {year} {2021})}\BibitemShut {NoStop}%
\bibitem [{\citenamefont {Bintrim}\ and\ \citenamefont {Berkelbach}(2022)}]{bintrim2022full}%
  \BibitemOpen
  \bibfield  {author} {\bibinfo {author} {\bibfnamefont {S~J}\ \bibnamefont {Bintrim}}\ and\ \bibinfo {author} {\bibfnamefont {T~C}\ \bibnamefont {Berkelbach}},\ }\bibfield  {title} {\enquote {\bibinfo {title} {Full-frequency dynamical {Bethe--Salpeter} equation without frequency and a study of double excitations},}\ }\href {https://doi.org/10.1063/5.0074434} {\bibfield  {journal} {\bibinfo  {journal} {The Journal of Chemical Physics}\ }\textbf {\bibinfo {volume} {156}},\ \bibinfo {pages} {044114} (\bibinfo {year} {2022})}\BibitemShut {NoStop}%
\bibitem [{\citenamefont {Scott}\ \emph {et~al.}(2023)\citenamefont {Scott}, \citenamefont {Backhouse},\ and\ \citenamefont {Booth}}]{scott2023moment}%
  \BibitemOpen
  \bibfield  {author} {\bibinfo {author} {\bibfnamefont {C~J~C}\ \bibnamefont {Scott}}, \bibinfo {author} {\bibfnamefont {O~J}\ \bibnamefont {Backhouse}}, \ and\ \bibinfo {author} {\bibfnamefont {G~H}\ \bibnamefont {Booth}},\ }\bibfield  {title} {\enquote {\bibinfo {title} {A moment-conserving reformulation of {$GW$} theory},}\ }\href {https://doi.org/10.1063/5.0143291} {\bibfield  {journal} {\bibinfo  {journal} {The Journal of Chemical Physics}\ }\textbf {\bibinfo {volume} {158}} (\bibinfo {year} {2023})}\BibitemShut {NoStop}%
\bibitem [{\citenamefont {Quintero-Monsebaiz}\ \emph {et~al.}(2022)\citenamefont {Quintero-Monsebaiz}, \citenamefont {Monino}, \citenamefont {Marie},\ and\ \citenamefont {Loos}}]{quintero2022connections}%
  \BibitemOpen
  \bibfield  {author} {\bibinfo {author} {\bibfnamefont {R}~\bibnamefont {Quintero-Monsebaiz}}, \bibinfo {author} {\bibfnamefont {E}~\bibnamefont {Monino}}, \bibinfo {author} {\bibfnamefont {A}~\bibnamefont {Marie}}, \ and\ \bibinfo {author} {\bibfnamefont {P-F}\ \bibnamefont {Loos}},\ }\bibfield  {title} {\enquote {\bibinfo {title} {Connections between many-body perturbation and coupled-cluster theories},}\ }\href {https://doi.org/10.1063/5.0130837} {\bibfield  {journal} {\bibinfo  {journal} {The Journal of Chemical Physics}\ }\textbf {\bibinfo {volume} {157}},\ \bibinfo {pages} {231102} (\bibinfo {year} {2022})}\BibitemShut {NoStop}%
\bibitem [{\citenamefont {Monino}\ and\ \citenamefont {Loos}(2023)}]{monino2023connections}%
  \BibitemOpen
  \bibfield  {author} {\bibinfo {author} {\bibfnamefont {E}~\bibnamefont {Monino}}\ and\ \bibinfo {author} {\bibfnamefont {P-F}\ \bibnamefont {Loos}},\ }\bibfield  {title} {\enquote {\bibinfo {title} {Connections and performances of {Green’s} function methods for charged and neutral excitations},}\ }\href {https://doi.org/10.1063/5.0159853} {\bibfield  {journal} {\bibinfo  {journal} {The Journal of Chemical Physics}\ }\textbf {\bibinfo {volume} {159}} (\bibinfo {year} {2023})}\BibitemShut {NoStop}%
\bibitem [{\citenamefont {T{\"o}lle}\ and\ \citenamefont {Chan}(2023)}]{tolle2023exact}%
  \BibitemOpen
  \bibfield  {author} {\bibinfo {author} {\bibfnamefont {J}~\bibnamefont {T{\"o}lle}}\ and\ \bibinfo {author} {\bibfnamefont {G~K-L}\ \bibnamefont {Chan}},\ }\bibfield  {title} {\enquote {\bibinfo {title} {Exact relationships between the {$GW$} approximation and equation-of-motion coupled-cluster theories through the quasi-boson formalism},}\ }\href {https://doi.org/10.1063/5.0139716} {\bibfield  {journal} {\bibinfo  {journal} {The Journal of Chemical Physics}\ }\textbf {\bibinfo {volume} {158}} (\bibinfo {year} {2023})}\BibitemShut {NoStop}%
\bibitem [{\citenamefont {Scuseria}\ \emph {et~al.}(2008)\citenamefont {Scuseria}, \citenamefont {Henderson},\ and\ \citenamefont {Sorensen}}]{scuseria2008ground}%
  \BibitemOpen
  \bibfield  {author} {\bibinfo {author} {\bibfnamefont {G~E}\ \bibnamefont {Scuseria}}, \bibinfo {author} {\bibfnamefont {T~M}\ \bibnamefont {Henderson}}, \ and\ \bibinfo {author} {\bibfnamefont {D~C}\ \bibnamefont {Sorensen}},\ }\bibfield  {title} {\enquote {\bibinfo {title} {The ground state correlation energy of the random phase approximation from a ring coupled cluster doubles approach},}\ }\href {https://doi.org/10.1063/1.3043729} {\bibfield  {journal} {\bibinfo  {journal} {The Journal of Chemical Physics}\ }\textbf {\bibinfo {volume} {129}},\ \bibinfo {pages} {231101} (\bibinfo {year} {2008})}\BibitemShut {NoStop}%
\bibitem [{\citenamefont {Berkelbach}(2018)}]{berkelbach2018communication}%
  \BibitemOpen
  \bibfield  {author} {\bibinfo {author} {\bibfnamefont {T~C}\ \bibnamefont {Berkelbach}},\ }\bibfield  {title} {\enquote {\bibinfo {title} {Communication: Random-phase approximation excitation energies from approximate equation-of-motion coupled-cluster doubles},}\ }\href {https://doi.org/10.1063/1.5032314} {\bibfield  {journal} {\bibinfo  {journal} {The Journal of Chemical Physics}\ }\textbf {\bibinfo {volume} {149}},\ \bibinfo {pages} {041103} (\bibinfo {year} {2018})}\BibitemShut {NoStop}%
\bibitem [{\citenamefont {Rishi}\ \emph {et~al.}(2020)\citenamefont {Rishi}, \citenamefont {Perera},\ and\ \citenamefont {Bartlett}}]{rishi2020route}%
  \BibitemOpen
  \bibfield  {author} {\bibinfo {author} {\bibfnamefont {V}~\bibnamefont {Rishi}}, \bibinfo {author} {\bibfnamefont {A}~\bibnamefont {Perera}}, \ and\ \bibinfo {author} {\bibfnamefont {R~J}\ \bibnamefont {Bartlett}},\ }\bibfield  {title} {\enquote {\bibinfo {title} {A route to improving {RPA} excitation energies through its connection to equation-of-motion coupled-cluster theory},}\ }\href {https://doi.org/10.1063/5.0023862} {\bibfield  {journal} {\bibinfo  {journal} {The Journal of Chemical Physics}\ }\textbf {\bibinfo {volume} {153}},\ \bibinfo {pages} {234101} (\bibinfo {year} {2020})}\BibitemShut {NoStop}%
\bibitem [{\citenamefont {Lange}\ and\ \citenamefont {Berkelbach}(2018)}]{lange2018relation}%
  \BibitemOpen
  \bibfield  {author} {\bibinfo {author} {\bibfnamefont {M~F}\ \bibnamefont {Lange}}\ and\ \bibinfo {author} {\bibfnamefont {T~C}\ \bibnamefont {Berkelbach}},\ }\bibfield  {title} {\enquote {\bibinfo {title} {On the relation between equation-of-motion coupled-cluster theory and the {$GW$} approximation},}\ }\href {https://doi.org/10.1021/acs.jctc.8b00455} {\bibfield  {journal} {\bibinfo  {journal} {Journal of Chemical Theory and Computation}\ }\textbf {\bibinfo {volume} {14}},\ \bibinfo {pages} {4224--4236} (\bibinfo {year} {2018})}\BibitemShut {NoStop}%
\bibitem [{\citenamefont {Galitskii}\ and\ \citenamefont {Migdal}(1958)}]{galitskii1958application}%
  \BibitemOpen
  \bibfield  {author} {\bibinfo {author} {\bibfnamefont {V~M}\ \bibnamefont {Galitskii}}\ and\ \bibinfo {author} {\bibfnamefont {A~B}\ \bibnamefont {Migdal}},\ }\bibfield  {title} {\enquote {\bibinfo {title} {Application of quantum field theory methods to the many-body problem},}\ }\href@noop {} {\bibfield  {journal} {\bibinfo  {journal} {Sov. Phys. JETP}\ }\textbf {\bibinfo {volume} {7}},\ \bibinfo {pages} {18} (\bibinfo {year} {1958})}\BibitemShut {NoStop}%
\bibitem [{\citenamefont {Hybertsen}\ and\ \citenamefont {Louie}(1986)}]{hybertsen1986electron}%
  \BibitemOpen
  \bibfield  {author} {\bibinfo {author} {\bibfnamefont {M~S}\ \bibnamefont {Hybertsen}}\ and\ \bibinfo {author} {\bibfnamefont {S~G}\ \bibnamefont {Louie}},\ }\bibfield  {title} {\enquote {\bibinfo {title} {Electron correlation in semiconductors and insulators: Band gaps and quasiparticle energies},}\ }\href {https://doi.org/10.1103/PhysRevB.34.5390} {\bibfield  {journal} {\bibinfo  {journal} {Physical Review B}\ }\textbf {\bibinfo {volume} {34}},\ \bibinfo {pages} {5390} (\bibinfo {year} {1986})}\BibitemShut {NoStop}%
\bibitem [{\citenamefont {Schirmer}\ \emph {et~al.}(1983)\citenamefont {Schirmer}, \citenamefont {Cederbaum},\ and\ \citenamefont {Walter}}]{schirmer1983new}%
  \BibitemOpen
  \bibfield  {author} {\bibinfo {author} {\bibfnamefont {J}~\bibnamefont {Schirmer}}, \bibinfo {author} {\bibfnamefont {L~S}\ \bibnamefont {Cederbaum}}, \ and\ \bibinfo {author} {\bibfnamefont {O}~\bibnamefont {Walter}},\ }\bibfield  {title} {\enquote {\bibinfo {title} {New approach to the one-particle {G}reen's function for finite {F}ermi systems},}\ }\href {https://doi.org/10.1103/PhysRevA.28.1237} {\bibfield  {journal} {\bibinfo  {journal} {Physical Review A}\ }\textbf {\bibinfo {volume} {28}},\ \bibinfo {pages} {1237} (\bibinfo {year} {1983})}\BibitemShut {NoStop}%
\bibitem [{\citenamefont {Caruso}\ \emph {et~al.}(2013)\citenamefont {Caruso}, \citenamefont {Rinke}, \citenamefont {Ren}, \citenamefont {Rubio},\ and\ \citenamefont {Scheffler}}]{caruso2013self}%
  \BibitemOpen
  \bibfield  {author} {\bibinfo {author} {\bibfnamefont {F}~\bibnamefont {Caruso}}, \bibinfo {author} {\bibfnamefont {P}~\bibnamefont {Rinke}}, \bibinfo {author} {\bibfnamefont {X}~\bibnamefont {Ren}}, \bibinfo {author} {\bibfnamefont {A}~\bibnamefont {Rubio}}, \ and\ \bibinfo {author} {\bibfnamefont {M}~\bibnamefont {Scheffler}},\ }\bibfield  {title} {\enquote {\bibinfo {title} {Self-consistent {$GW$}: All-electron implementation with localized basis functions},}\ }\href {https://doi.org/10.1103/PhysRevB.88.075105} {\bibfield  {journal} {\bibinfo  {journal} {Physical Review B}\ }\textbf {\bibinfo {volume} {88}},\ \bibinfo {pages} {075105} (\bibinfo {year} {2013})}\BibitemShut {NoStop}%
\bibitem [{\citenamefont {Raimondi}\ and\ \citenamefont {Barbieri}(2018)}]{raimondi2018algebraic}%
  \BibitemOpen
  \bibfield  {author} {\bibinfo {author} {\bibfnamefont {F}~\bibnamefont {Raimondi}}\ and\ \bibinfo {author} {\bibfnamefont {C}~\bibnamefont {Barbieri}},\ }\bibfield  {title} {\enquote {\bibinfo {title} {Algebraic diagrammatic construction formalism with three-body interactions},}\ }\href {https://doi.org/10.1103/PhysRevC.97.054308} {\bibfield  {journal} {\bibinfo  {journal} {Physical Review C}\ }\textbf {\bibinfo {volume} {97}},\ \bibinfo {pages} {054308} (\bibinfo {year} {2018})}\BibitemShut {NoStop}%
\bibitem [{\citenamefont {Schirmer}(2018)}]{schirmer2018many}%
  \BibitemOpen
  \bibfield  {author} {\bibinfo {author} {\bibfnamefont {J}~\bibnamefont {Schirmer}},\ }\href@noop {} {\emph {\bibinfo {title} {Many-body methods for atoms, molecules and clusters}}},\ Vol.~\bibinfo {volume} {94}\ (\bibinfo  {publisher} {Springer},\ \bibinfo {year} {2018})\BibitemShut {NoStop}%
\bibitem [{\citenamefont {Coveney}\ and\ \citenamefont {Tew}(2023)}]{coveney2023regularized}%
  \BibitemOpen
  \bibfield  {author} {\bibinfo {author} {\bibfnamefont {C~J~N}\ \bibnamefont {Coveney}}\ and\ \bibinfo {author} {\bibfnamefont {D~P}\ \bibnamefont {Tew}},\ }\bibfield  {title} {\enquote {\bibinfo {title} {A regularized second-order correlation method from green’s function theory},}\ }\href {https://doi.org/10.1021/acs.jctc.3c00246} {\bibfield  {journal} {\bibinfo  {journal} {Journal of Chemical Theory and Computation}\ }\textbf {\bibinfo {volume} {19}},\ \bibinfo {pages} {3915--3928} (\bibinfo {year} {2023})}\BibitemShut {NoStop}%
\bibitem [{\citenamefont {Backhouse}\ and\ \citenamefont {Booth}(2020)}]{backhouse2020efficient}%
  \BibitemOpen
  \bibfield  {author} {\bibinfo {author} {\bibfnamefont {O~J}\ \bibnamefont {Backhouse}}\ and\ \bibinfo {author} {\bibfnamefont {G~H}\ \bibnamefont {Booth}},\ }\bibfield  {title} {\enquote {\bibinfo {title} {Efficient excitations and spectra within a perturbative renormalization approach},}\ }\href {https://doi.org/10.1021/acs.jctc.0c00701} {\bibfield  {journal} {\bibinfo  {journal} {Journal of Chemical Theory and Computation}\ }\textbf {\bibinfo {volume} {16}},\ \bibinfo {pages} {6294--6304} (\bibinfo {year} {2020})}\BibitemShut {NoStop}%
\bibitem [{\citenamefont {Scuseria}(1995)}]{scuseria1995connections}%
  \BibitemOpen
  \bibfield  {author} {\bibinfo {author} {\bibfnamefont {G~E}\ \bibnamefont {Scuseria}},\ }\bibfield  {title} {\enquote {\bibinfo {title} {On the connections between {Brueckner}--coupled-cluster, density-dependent {Hartree--Fock}, and {Density Functional Theory}},}\ }\href {https://doi.org/10.1002/qua.560550211} {\bibfield  {journal} {\bibinfo  {journal} {International Journal of Quantum Chemistry}\ }\textbf {\bibinfo {volume} {55}},\ \bibinfo {pages} {165--171} (\bibinfo {year} {1995})}\BibitemShut {NoStop}%
\bibitem [{\citenamefont {Helgaker}\ \emph {et~al.}(2013)\citenamefont {Helgaker}, \citenamefont {Jorgensen},\ and\ \citenamefont {Olsen}}]{helgaker2013molecular}%
  \BibitemOpen
  \bibfield  {author} {\bibinfo {author} {\bibfnamefont {T}~\bibnamefont {Helgaker}}, \bibinfo {author} {\bibfnamefont {P}~\bibnamefont {Jorgensen}}, \ and\ \bibinfo {author} {\bibfnamefont {J}~\bibnamefont {Olsen}},\ }\href@noop {} {\emph {\bibinfo {title} {Molecular electronic-structure theory}}}\ (\bibinfo  {publisher} {John Wiley \& Sons},\ \bibinfo {year} {2013})\BibitemShut {NoStop}%
\bibitem [{\citenamefont {Shavitt}\ and\ \citenamefont {Bartlett}(2009)}]{shavitt2009many}%
  \BibitemOpen
  \bibfield  {author} {\bibinfo {author} {\bibfnamefont {I}~\bibnamefont {Shavitt}}\ and\ \bibinfo {author} {\bibfnamefont {R~J}\ \bibnamefont {Bartlett}},\ }\href@noop {} {\emph {\bibinfo {title} {Many-body methods in chemistry and physics: MBPT and coupled-cluster theory}}}\ (\bibinfo  {publisher} {Cambridge University Press},\ \bibinfo {year} {2009})\BibitemShut {NoStop}%
\bibitem [{\citenamefont {Gauss}\ and\ \citenamefont {Stanton}(1995)}]{gauss1995coupled}%
  \BibitemOpen
  \bibfield  {author} {\bibinfo {author} {\bibfnamefont {J}~\bibnamefont {Gauss}}\ and\ \bibinfo {author} {\bibfnamefont {J~F}\ \bibnamefont {Stanton}},\ }\bibfield  {title} {\enquote {\bibinfo {title} {Coupled-cluster calculations of nuclear magnetic resonance chemical shifts},}\ }\href {https://doi.org/10.1063/1.470240} {\bibfield  {journal} {\bibinfo  {journal} {The Journal of Chemical Physics}\ }\textbf {\bibinfo {volume} {103}},\ \bibinfo {pages} {3561--3577} (\bibinfo {year} {1995})}\BibitemShut {NoStop}%
\bibitem [{\citenamefont {Hirata}\ \emph {et~al.}(2024)\citenamefont {Hirata}, \citenamefont {Grabowski}, \citenamefont {Ortiz},\ and\ \citenamefont {Bartlett}}]{hirata2024nonconvergence}%
  \BibitemOpen
  \bibfield  {author} {\bibinfo {author} {\bibfnamefont {S}~\bibnamefont {Hirata}}, \bibinfo {author} {\bibfnamefont {I}~\bibnamefont {Grabowski}}, \bibinfo {author} {\bibfnamefont {J~V}\ \bibnamefont {Ortiz}}, \ and\ \bibinfo {author} {\bibfnamefont {R~J}\ \bibnamefont {Bartlett}},\ }\bibfield  {title} {\enquote {\bibinfo {title} {Nonconvergence of the {Feynman-Dyson} diagrammatic perturbation expansion of propagators},}\ }\href {https://doi.org/10.1103/PhysRevA.109.052220} {\bibfield  {journal} {\bibinfo  {journal} {Physical Review A}\ }\textbf {\bibinfo {volume} {109}},\ \bibinfo {pages} {052220} (\bibinfo {year} {2024})}\BibitemShut {NoStop}%
\bibitem [{\citenamefont {Scuseria}\ \emph {et~al.}(2013)\citenamefont {Scuseria}, \citenamefont {Henderson},\ and\ \citenamefont {Bulik}}]{scuseria2013particle}%
  \BibitemOpen
  \bibfield  {author} {\bibinfo {author} {\bibfnamefont {G~E}\ \bibnamefont {Scuseria}}, \bibinfo {author} {\bibfnamefont {T~M}\ \bibnamefont {Henderson}}, \ and\ \bibinfo {author} {\bibfnamefont {I~W}\ \bibnamefont {Bulik}},\ }\bibfield  {title} {\enquote {\bibinfo {title} {Particle-particle and quasiparticle random phase approximations: Connections to coupled-cluster theory},}\ }\href {https://doi.org/10.1063/1.4820557} {\bibfield  {journal} {\bibinfo  {journal} {The Journal of Chemical Physics}\ }\textbf {\bibinfo {volume} {139}},\ \bibinfo {pages} {104113} (\bibinfo {year} {2013})}\BibitemShut {NoStop}%
\bibitem [{\citenamefont {Rohlfing}\ and\ \citenamefont {Louie}(2000)}]{rohlfing2000electron}%
  \BibitemOpen
  \bibfield  {author} {\bibinfo {author} {\bibfnamefont {M}~\bibnamefont {Rohlfing}}\ and\ \bibinfo {author} {\bibfnamefont {S~G}\ \bibnamefont {Louie}},\ }\bibfield  {title} {\enquote {\bibinfo {title} {Electron-hole excitations and optical spectra from first principles},}\ }\href {https://doi.org/10.1103/PhysRevB.62.4927} {\bibfield  {journal} {\bibinfo  {journal} {Physical Review B}\ }\textbf {\bibinfo {volume} {62}},\ \bibinfo {pages} {4927} (\bibinfo {year} {2000})}\BibitemShut {NoStop}%
\bibitem [{\citenamefont {Ring}\ and\ \citenamefont {Schuck}(2004)}]{ring2004nuclear}%
  \BibitemOpen
  \bibfield  {author} {\bibinfo {author} {\bibfnamefont {P}~\bibnamefont {Ring}}\ and\ \bibinfo {author} {\bibfnamefont {P}~\bibnamefont {Schuck}},\ }\href@noop {} {\emph {\bibinfo {title} {The nuclear many-body problem}}}\ (\bibinfo  {publisher} {Springer Science \& Business Media},\ \bibinfo {year} {2004})\BibitemShut {NoStop}%
\bibitem [{\citenamefont {Schindlmayr}\ and\ \citenamefont {Godby}(1998)}]{schindlmayr1998systematic}%
  \BibitemOpen
  \bibfield  {author} {\bibinfo {author} {\bibfnamefont {A}~\bibnamefont {Schindlmayr}}\ and\ \bibinfo {author} {\bibfnamefont {R~W}\ \bibnamefont {Godby}},\ }\bibfield  {title} {\enquote {\bibinfo {title} {Systematic vertex corrections through iterative solution of {Hedin's} equations beyond the {$GW$} approximation},}\ }\href {https://doi.org/10.1103/PhysRevLett.80.1702} {\bibfield  {journal} {\bibinfo  {journal} {Physical Review Letters}\ }\textbf {\bibinfo {volume} {80}},\ \bibinfo {pages} {1702} (\bibinfo {year} {1998})}\BibitemShut {NoStop}%
\bibitem [{\citenamefont {Molinari}(2005)}]{molinari2005hedin}%
  \BibitemOpen
  \bibfield  {author} {\bibinfo {author} {\bibfnamefont {L~G}\ \bibnamefont {Molinari}},\ }\bibfield  {title} {\enquote {\bibinfo {title} {Hedin’s equations and enumeration of {Feynman} diagrams},}\ }\href {https://doi.org/10.1103/PhysRevB.71.113102} {\bibfield  {journal} {\bibinfo  {journal} {Physical Review B}\ }\textbf {\bibinfo {volume} {71}},\ \bibinfo {pages} {113102} (\bibinfo {year} {2005})}\BibitemShut {NoStop}%
\bibitem [{\citenamefont {Ness}\ \emph {et~al.}(2011)\citenamefont {Ness}, \citenamefont {Dash}, \citenamefont {Stankovski},\ and\ \citenamefont {Godby}}]{ness2011g}%
  \BibitemOpen
  \bibfield  {author} {\bibinfo {author} {\bibfnamefont {H}~\bibnamefont {Ness}}, \bibinfo {author} {\bibfnamefont {L~K}\ \bibnamefont {Dash}}, \bibinfo {author} {\bibfnamefont {M}~\bibnamefont {Stankovski}}, \ and\ \bibinfo {author} {\bibfnamefont {R~W}\ \bibnamefont {Godby}},\ }\bibfield  {title} {\enquote {\bibinfo {title} {{$GW$} approximations and vertex corrections on the {Keldysh} time-loop contour: Application for model systems at equilibrium},}\ }\href {https://doi.org/10.1103/PhysRevB.84.195114} {\bibfield  {journal} {\bibinfo  {journal} {Physical Review B}\ }\textbf {\bibinfo {volume} {84}},\ \bibinfo {pages} {195114} (\bibinfo {year} {2011})}\BibitemShut {NoStop}%
\bibitem [{\citenamefont {Lani}\ \emph {et~al.}(2012)\citenamefont {Lani}, \citenamefont {Romaniello},\ and\ \citenamefont {Reining}}]{lani2012approximations}%
  \BibitemOpen
  \bibfield  {author} {\bibinfo {author} {\bibfnamefont {G}~\bibnamefont {Lani}}, \bibinfo {author} {\bibfnamefont {P}~\bibnamefont {Romaniello}}, \ and\ \bibinfo {author} {\bibfnamefont {L}~\bibnamefont {Reining}},\ }\bibfield  {title} {\enquote {\bibinfo {title} {Approximations for many-body {Green's} functions: insights from the fundamental equations},}\ }\href {https://iopscience.iop.org/article/10.1088/1367-2630/14/1/013056/meta} {\bibfield  {journal} {\bibinfo  {journal} {New Journal of Physics}\ }\textbf {\bibinfo {volume} {14}},\ \bibinfo {pages} {013056} (\bibinfo {year} {2012})}\BibitemShut {NoStop}%
\bibitem [{\citenamefont {Kutepov}(2017)}]{kutepov2017self}%
  \BibitemOpen
  \bibfield  {author} {\bibinfo {author} {\bibfnamefont {A~L}\ \bibnamefont {Kutepov}},\ }\bibfield  {title} {\enquote {\bibinfo {title} {Self-consistent solution of {Hedin's} equations: semiconductors and insulators},}\ }\href {https://doi.org/10.1103/PhysRevB.95.195120} {\bibfield  {journal} {\bibinfo  {journal} {Physical Review B}\ }\textbf {\bibinfo {volume} {95}},\ \bibinfo {pages} {195120} (\bibinfo {year} {2017})}\BibitemShut {NoStop}%
\bibitem [{\citenamefont {Harkov}\ \emph {et~al.}(2021)\citenamefont {Harkov}, \citenamefont {Lichtenstein},\ and\ \citenamefont {Krien}}]{harkov2021parametrizations}%
  \BibitemOpen
  \bibfield  {author} {\bibinfo {author} {\bibfnamefont {V}~\bibnamefont {Harkov}}, \bibinfo {author} {\bibfnamefont {A~I}\ \bibnamefont {Lichtenstein}}, \ and\ \bibinfo {author} {\bibfnamefont {F}~\bibnamefont {Krien}},\ }\bibfield  {title} {\enquote {\bibinfo {title} {Parametrizations of local vertex corrections from weak to strong coupling: Importance of the {Hedin} three-leg vertex},}\ }\href {https://doi.org/10.1103/PhysRevB.104.125141} {\bibfield  {journal} {\bibinfo  {journal} {Physical Review B}\ }\textbf {\bibinfo {volume} {104}},\ \bibinfo {pages} {125141} (\bibinfo {year} {2021})}\BibitemShut {NoStop}%
\bibitem [{\citenamefont {Kutepov}(2022)}]{kutepov2022full}%
  \BibitemOpen
  \bibfield  {author} {\bibinfo {author} {\bibfnamefont {A~L}\ \bibnamefont {Kutepov}},\ }\bibfield  {title} {\enquote {\bibinfo {title} {Full versus quasiparticle self-consistency in vertex-corrected {$GW$} approaches},}\ }\href {https://doi.org/10.1103/PhysRevB.105.045124} {\bibfield  {journal} {\bibinfo  {journal} {Physical Review B}\ }\textbf {\bibinfo {volume} {105}},\ \bibinfo {pages} {045124} (\bibinfo {year} {2022})}\BibitemShut {NoStop}%
\bibitem [{\citenamefont {Mejuto-Zaera}\ and\ \citenamefont {Vl{\v{c}}ek}(2022)}]{mejuto2022self}%
  \BibitemOpen
  \bibfield  {author} {\bibinfo {author} {\bibfnamefont {C}~\bibnamefont {Mejuto-Zaera}}\ and\ \bibinfo {author} {\bibfnamefont {V}~\bibnamefont {Vl{\v{c}}ek}},\ }\bibfield  {title} {\enquote {\bibinfo {title} {Self-consistency in {$GW\Gamma$} formalism leading to quasiparticle-quasiparticle couplings},}\ }\href {https://doi.org/10.1103/PhysRevB.106.165129} {\bibfield  {journal} {\bibinfo  {journal} {Physical Review B}\ }\textbf {\bibinfo {volume} {106}},\ \bibinfo {pages} {165129} (\bibinfo {year} {2022})}\BibitemShut {NoStop}%
\bibitem [{\citenamefont {Riva}\ \emph {et~al.}(2023)\citenamefont {Riva}, \citenamefont {Romaniello},\ and\ \citenamefont {Berger}}]{riva2023multichannel}%
  \BibitemOpen
  \bibfield  {author} {\bibinfo {author} {\bibfnamefont {G}~\bibnamefont {Riva}}, \bibinfo {author} {\bibfnamefont {P}~\bibnamefont {Romaniello}}, \ and\ \bibinfo {author} {\bibfnamefont {A~J}\ \bibnamefont {Berger}},\ }\bibfield  {title} {\enquote {\bibinfo {title} {Multichannel {Dyson} equation: Coupling many-body {Green’s} functions},}\ }\href {https://doi.org/10.1103/PhysRevLett.131.216401} {\bibfield  {journal} {\bibinfo  {journal} {Physical Review Letters}\ }\textbf {\bibinfo {volume} {131}},\ \bibinfo {pages} {216401} (\bibinfo {year} {2023})}\BibitemShut {NoStop}%
\bibitem [{\citenamefont {L{\"o}wdin}(1982)}]{lowdin1982partitioning}%
  \BibitemOpen
  \bibfield  {author} {\bibinfo {author} {\bibfnamefont {P-O}\ \bibnamefont {L{\"o}wdin}},\ }\bibfield  {title} {\enquote {\bibinfo {title} {Partitioning technique, perturbation theory, and rational approximations},}\ }\href {https://doi.org/10.1002/qua.560210105} {\bibfield  {journal} {\bibinfo  {journal} {International Journal of Quantum Chemistry}\ }\textbf {\bibinfo {volume} {21}},\ \bibinfo {pages} {69--92} (\bibinfo {year} {1982})}\BibitemShut {NoStop}%
\end{thebibliography}%

\end{document}